\documentclass[aps,prd,nofootinbib,floatfix,superscriptaddress,showpacs,showkeys, eqsecnum,11pt]{revtex4-1}
\usepackage{amsmath,amssymb,amsthm,bm,bbm,color}
\usepackage{graphicx,psfrag,subfigure,rotating}

\begin{document}

\title{Intersections of thick Center Vortices, Dirac Eigenmodes and\\ Fractional Topological Charge in $SU(2)$ Lattice Gauge Theory}

\author{R. H\"ollwieser}
\affiliation{Atomic Institute, Vienna University of Technology, Wiedner Hauptstr.\ 8-10, A-1040 Vienna, Austria}

\author{M. Faber}
\affiliation{Atomic Institute, Vienna University of Technology, Wiedner Hauptstr.\ 8-10, A-1040 Vienna, Austria}

\author{U.M. Heller}
\affiliation{American Physical Society, One Research Road, Ridge, NY 11961, USA}

\date{\today}

\begin{abstract}
Intersections of thick, plane $SU(2)$ center vortices are characterized by the topological charge $|Q|=1/2$. We compare such intersections with the
distribution of zeromodes of the Dirac operator in the fundamental and adjoint representation using both the overlap and asqtad staggered fermion
formulations in $SU(2)$ lattice gauge theory. We analyze configurations with
four intersections and find that the probability density distribution of
fundamental zeromodes in the intersection plane differs significantly from the one obtained analytically in~\cite{Reinhardt:2002cm}. The Dirac
eigenmodes are clearly sensitive to the traces of the Polyakov (Wilson) lines
and do not exactly locate topological charge contributions. Although, the
adjoint Dirac operator is able to produce zeromodes for configurations with
topological charge $|Q|=1/2$, they do not locate single vortex
intersections, as we prove by forming arbitrary linear combinations of these
zeromodes - their scalar density peaks at least at two intersection points.
With pairs of thin and thick vortices we realize a situation similar to
configurations with topological charge $|Q|=1/2$. For such configurations the zeromodes do not localize in the regions of fractional topological charge contribution but spread over the whole lattice, avoiding regions of negative traces of Polyakov lines. This sensitivity to Polyakov lines we also confirm for single vortex-pairs, {\it i.e.}, configurations with nontrivial Polyakov loops but without topological charge.
\end{abstract}

\maketitle

\section{Introduction}
 Quantum chromodynamics (QCD) at low energies is dominated by the
 non-perturbative phenomena of quark confinement and spontaneous chiral symmetry
 breaking (SCSB). Presently, a rigorous treatment of them is only possible
 in a lattice regularization. Many of the important features of non-abelian
 gauge theories are already present in $SU(2)$, which simplifies theoretical and numerical calculations.

The non-perturbative vacuum can be characterized by various kinds of topological gauge field excitations. An established theory of SCSB relies on instantons~\cite{Belavin:1975fg,Actor:1979in,'tHooft:1976fv,Bernard:1979qt}
, which are localized in space-time and carry a topological charge of modulus one. Due to the Atiyah-Singer index theorem~\cite{Atiyah:1971rm,Schwarz:1977az,Brown:1977bj,Narayanan:1994gw}, a zeromode of the Dirac operator arises, which is concentrated at the instanton core. In the instanton liquid model~\cite{Ilgenfritz:1980vj,Diakonov:1984vw} overlapping would-be zeromodes split into low lying non-zeromodes which create the chiral condensate.

On the other hand there is plenty of evidence~\cite{'tHooft:1977hy,Vinciarelli:1978kp,Yoneya:1978dt,Cornwall:1979hz,Mack:1978rq,Nielsen:1979xu} for the explanation of confinement by center vortices, closed, quantized magnetic flux tubes with values in the center of the gauge group. These properties are the key ingredients in the vortex model of confinement, which is theoretically appealing and was also confirmed by a multitude of numerical calculations~\cite{DelDebbio:1996mh,Kovacs:1998xm}. Lattice simulations indicate that vortices may be responsible for topological charge and SCSB as well~\cite{deForcrand:1999ms,Alexandrou:1999vx,Engelhardt:2002qs,Hollwieser:2008tq}, and thus unify all non-perturbative phenomena in a common framework. However, to date the potential physical mechanism for symmetry breaking through vortices is still unclear. A similar picture to the instanton liquid model exists insofar as lumps of topological charge arise at the intersection and writhing points of vortices~\cite{Reinhardt:2000ck, Reinhardt:2001kf}. 

The present numerical investigation concentrates on the topological charge contributions of vortex intersections and their localization by Dirac zeromodes. All measurements were performed on hyper-cubic lattices of even sizes from $12^4$ up to $22^4$-lattices. We start with the construction of thick planar vortex configurations and derive the topological charge contribution of their intersections. Then we introduce the lattice index theorem for various fermion realizations and representations in more detail. 

Using the overlap and asqtad staggered Dirac operator, we compute
fundamental zeromodes in the background of four vortex intersections. By
visualizing the probability density, we compare the distribution of the
eigenmode density with the position of the vortices and the topological
charge density created by intersection points. For interpreting the results, we
refer to existing analytical calculations of the zeromodes for flat
vortices with gauge potentials living in the Cartan subalgebra of a
$SU(2)$ gauge group~\cite{Reinhardt:2002cm}. A comparison reveals
systematic discrepancies, which we tentatively attribute to differences in the
values of the Polyakov loops between the analytical~\cite{Reinhardt:2002cm} and
our numerical calculations. We argue that configurations with the same field
strength, but different Polyakov loops do not give rise to the same
eigenmode density. 
Comments on the physical significance of this observation follow further below.

Furthermore, we calculate the Dirac eigenmodes in the adjoint fermion
representation. Adjoint fermions are of special interest with respect to the
constituents of the QCD vacuum with fractional topological charge. For
configurations with total topological charge $|Q|=1/2$ no fundamental
zeromode would necessarily be produced, however, adjoint fermions are able to create a zeromode. Edwards et al.\ presented in~\cite{Heller:1998cf} some evidence for fractional topological charge on the lattice. Garc\'ia-P\'erez et al.~\cite{GarciaPerez:2007ne}, however, associated this to lattice artifacts, {\it i.e.}, topological objects of size of the order of the lattice spacing. 

Vortex intersections are examples of fractional topological charge contributions $|Q|=1/2$, which can be related to merons~\cite{Reinhardt:2001hb} and calorons~\cite{Bruckmann:2009pa}. Nevertheless, on lattice configurations with periodic (untwisted) boundary conditions, no single vortex intersection can be realized. Therefore we tried to separate adjoint zeromodes on periodic lattices with four vortex intersections in order to find linear combinations of zeromodes which detect one vortex intersection only. We did not find such linear combinations, the scalar density of the zeromodes peaks at least at two intersection points.

Finally we combine "thin" and "thick" vortex sheets, which simulate in some
sense twisted boundary conditions. In fact, "thin" vortices are not
recognized by adjoint fermions. If we intersect such vortex pairs adjoint
fermions detect a single vortex intersection only and the adjoint index
theorem signals the topological charge $Q=1/2$. Nevertheless, the zeromodes do not seem to localize the vortex intersection but rather extend over the whole lattice, avoiding regions with nonvanishing topological charge density and negative Polyakov lines. 


\section{Plane Vortices}\label{sec:planevort}
\begin{figure}[htb]
\centering
\psfrag{z1}{\scriptsize $z_1$}
\psfrag{z2}{\scriptsize $z_2$}
\psfrag{2d}{\scriptsize $2d$}
\psfrag{1}{\scriptsize $1$}
\psfrag{0}{\scriptsize $0$}
\psfrag{p}{\scriptsize $\pi$}
\psfrag{2p}{\scriptsize $2\pi$}
\psfrag{12}{\scriptsize $N_z$}
\psfrag{f}{$\phi_2$}
\psfrag{f1}{$\phi_1$}
\psfrag{z}{$z$}
a)\includegraphics[width=.4\linewidth]{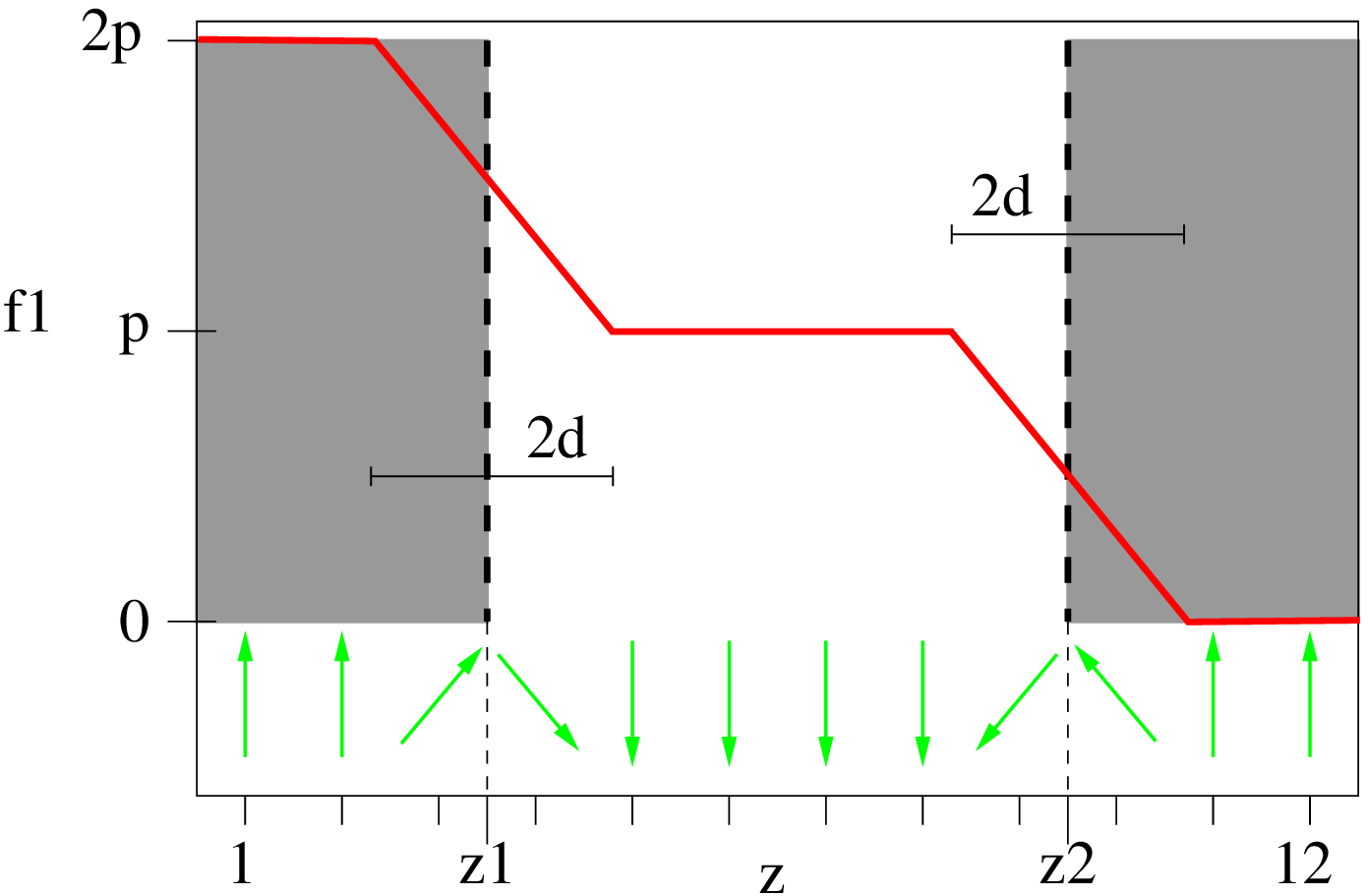}\hspace{5mm}b)\includegraphics[width=.4\linewidth]{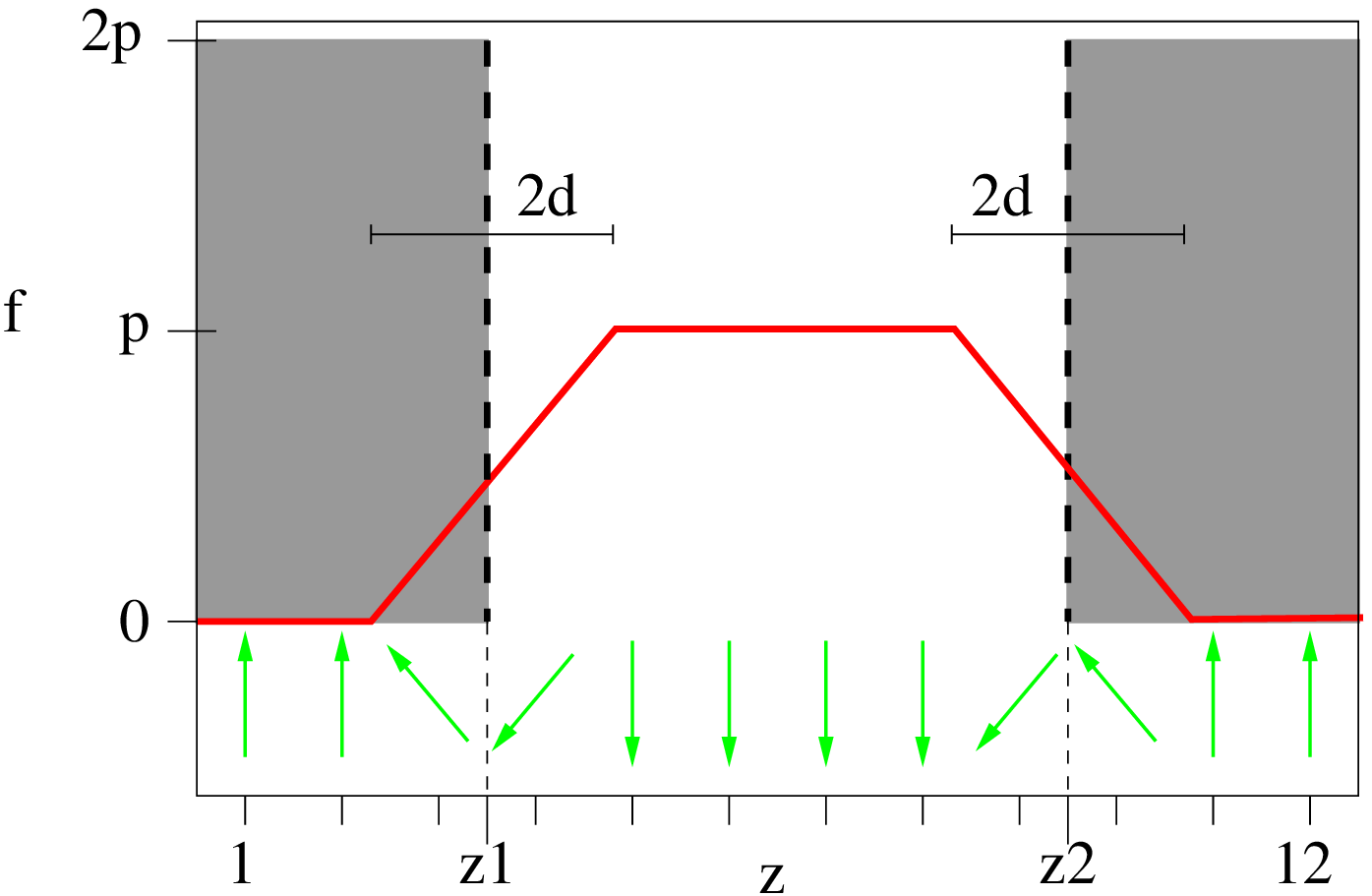}
\caption{The link angle a) $\phi_1$ of a parallel and b) $\phi_2$ of an anti-parallel vortex pair. The arrows rotate counterclockwise with increasing $\phi_i$. The vertical dashed lines indicate the positions of the P-vortices. In the shaded areas the links have positive, otherwise negative trace.}
\label{fig:phis}
\end{figure}

We investigate planar vortices parallel to two of the coordinate axes in
$SU(2)$ lattice gauge theory. Since we use periodic (untwisted) boundary
conditions for the links, vortices occur in pairs of parallel sheets, each of
which is closed by virtue of the lattice periodicity. We use two different
orientations of vortex sheets, $xy$- and $zt$-planes with nontrivial links
within the vortex thickness. These links vary in the $\sigma_3$ subgroup of
$SU(2)$, $U_\mu=\exp(i \phi \sigma_3)$. For $xy$-vortices $\mu=t$ links are
nontrivial in one $t$-slice only, for $zt$-vortices we have nontrivial
$y$-links in one $y$-slice. Since the $U(1)$ subgroup remains unchanged, the
direction of the flux and the orientation of the vortex are determined by the
gradient of the angle $\phi$, which we choose as a linear function of the
coordinate perpendicular to the vortex. Upon traversing a vortex sheet, the
angle $\phi$ increases or decreases by $\pi$ within a finite thickness $2d$ of
the vortex, see Fig.~\ref{fig:phis}. Center projection leads to a (thin)
P-vortex at half the thickness ($d$)~\cite{DelDebbio:1996mh}. We distinguish
parallel and anti-parallel vortices, {\it i.e.}, vortex sheets with the same resp. opposite orientation. The angle $\phi_i$ first increases from $0$ to $\pi$, then $\phi_1$ increases to $2\pi$ and $\phi_2$ returns to $0$. For an $xy$-vortex with vortex sheets at $z_1$ and $z_2$ the $t$-links in one $t$-slice vary with the angle
\begin{equation}
  \phi_1(z) = \begin{cases}     2\pi \\ \pi\left[ 2-\frac{z-(z_1-d)}{2d}\right] \\ 
                           \pi \\ \pi\left[ 1-\frac{z-(z_2-d)}{2d}\right] \\ 
                          0 \end{cases} \ldots
  \phi_2(z) = \begin{cases}     0 & 0 < z \leq z_1-d \\ 
                \frac{\pi}{2d}[z-(z_1-d)] & z_1-d < z \leq z_1+d \\ 
                \pi & z_1+d < z \leq z_2-d \\ 
                \pi\left[ 1-\frac{z-(z_2-d)}{2d}\right] & z_2-d < z \leq z_2+d \\ 
                0 & z_2+d < z \leq N \end{cases} \label{eq:phi-pl0}
\end{equation}
Since the gradient of $\phi_1$ ($\phi_2$) points in the same (opposite) direction at the two vortex sheets of a pair, their fluxes are (anti-) parallel and the total flux through the $zt$-plane is therefore $2\pi$ (zero).

Next we consider these thick, planar vortices intersecting orthogonally. As shown in~\cite{Engelhardt:1999xw}, each intersection carries a topological charge with modulus $|Q|=1/2$, whose sign depends on the relative orientation of the vortex fluxes. The plaquette definition simply discretizes the continuum (Minkowski) expression of the Pontryagin index to a lattice (Euclidean) version of the topological charge definition:
\begin{gather}
  Q = - \frac{1}{16\pi^2} \int d^4x \, \mbox{tr}[\tilde{\cal F}_{\mu\nu} {\cal F}_{\mu\nu} ] = - \frac{1}{32\pi^2} \int d^4x \, \epsilon_{\mu\nu\alpha\beta} \mbox{tr}[{\cal F}_{\alpha\beta} {\cal F}_{\mu\nu} ] = \frac{1}{4\pi^2} \int d^4x \, \vec E \cdot \vec B \label{eq:qlatq}
\end{gather}
We can drop color indices since all our links belong to the
$\sigma_3$-generated $U(1)$ subgroup of $SU(2)$. Our $xy$-vortices have only
nontrivial $zt$-plaquettes, {\it i.e.}, an electric field $E_z$, while
$zt$-vortices bear nontrivial $xy$-plaquettes corresponding to a magnetic
field $B_z$. The topological charge is then proportional to $E_zB_z$, hence
parallel crossings give $Q=1/2$ and anti-parallel crossings give $Q=-1/2$.
However, due to the finite lattice spacing (or the finite thickness of
vortices in units of the lattice spacing) the numerical absolute values of the lattice charge are slightly smaller than the continuum ones.

\begin{figure}[htb]
\psfrag{x}{$x$}
\psfrag{y}{$y$}
\psfrag{z}{$z$}
\psfrag{0}{\small $0$}
\psfrag{+Q}{\small $+Q$}
\psfrag{-Q}{\small $-Q$}
\begin{tabular}{ccc}
Parallel Vortices & Geometry & Anti-parallel Vortices\\
\includegraphics[keepaspectratio,width=0.33\textwidth]{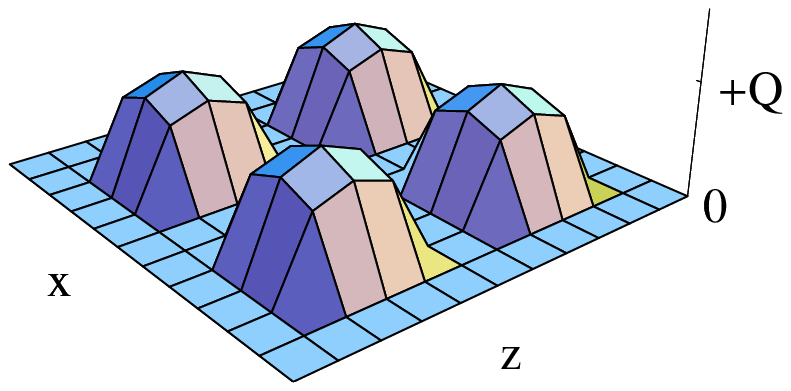} & 
\includegraphics[keepaspectratio,width=0.24\textwidth]{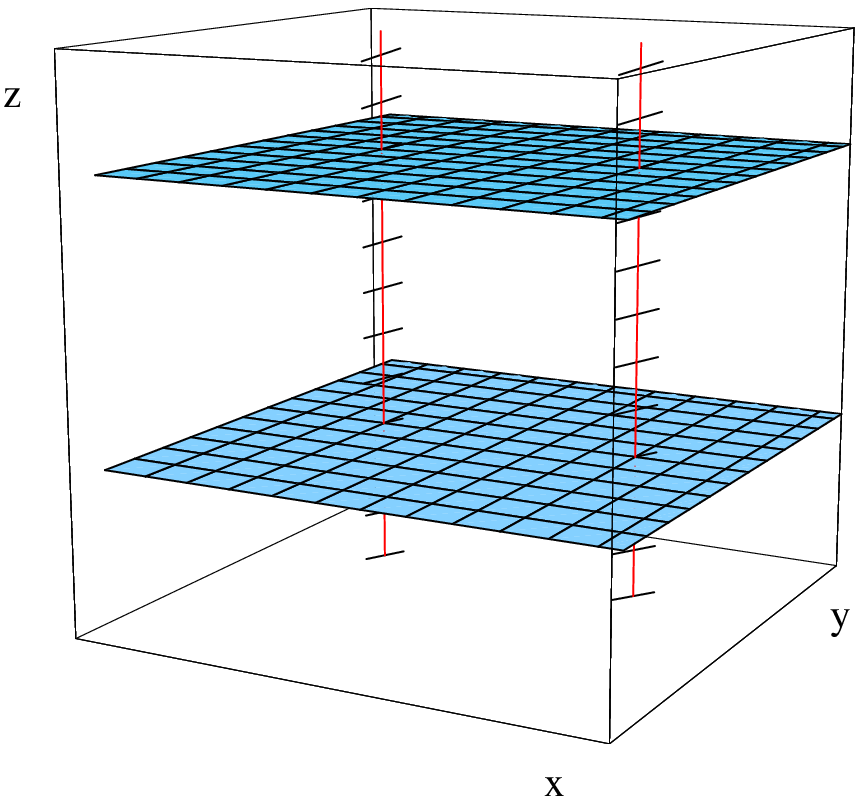} &
\;\includegraphics[keepaspectratio,width=0.33\textwidth]{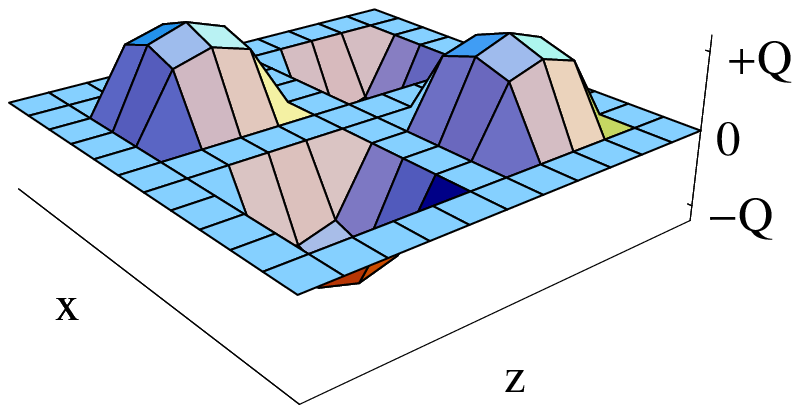}
\end{tabular}
\caption{A 3-dimensional section (hyperplane) in $xyz$-direction of a $12^4$-lattice at time $t=6$ (center). The horizontal planes are the $xy$-vortices, which exist only at this time. The vertical lines are the $zt$-vortices, which continue over the whole time axis. The ticks protruding from the vertical lines extend in time direction. The vortices intersect in four points of the $y=t=6$ - plane, giving topological charge $Q=2$ for parallel vortices (lhs) or $Q=0$ for anti-parallel vortices (rhs).}
\label{fig:vortcross3d}
\end{figure}

As mentioned above, the lattice periodicity forbids single vortex sheets and therefore we get at least four intersections for two vortex pairs, summing up to an even valued topological charge. 
We combine $xy$ and $zt$-vortices in central $y$- and $t$-slices with vortex centers at $x_{1,2}$ resp. $z_{1,2}$ located symmetrically around the lattice center and varying vortex thickness $d$. In Fig.~\ref{fig:vortcross3d} we present a 3-dimensional view of the intersecting vortices on a $12^4$-lattice together with topological charge distributions in the $xz$-plane at ($y=6,t=6$), which is the intersection plane. Further density plots of Dirac eigenmodes or topological charge, as well as Polyakov loop distributions will all be plotted in this plane, except for (specially mentioned) orthogonal plots in order to analyze localization properties. The four intersection points of two parallel vortices all carry topological charge $Q=+1/2$ whereas for two anti-parallel vortices and parallel-anti-parallel vortex combinations we get two intersections with $Q=+1/2$ and two with $Q=-1/2$.  

\section{Fermionic zeromodes of the overlap and asqtad staggered Dirac operator for intersecting center vortex fields}

We test the lattice index theorem and analyze the scalar density
$\rho(x)=\psi^\dag\psi(x)$ of fermionic zeromodes $\psi$ in the background of
intersecting plane vortices. As described in~\cite{Hollwieser:2010mj} the
improved staggered operator also produces eigenmodes which can clearly be
identified as zeromodes and all results in this paper show perfect agreement
between the two fermion realizations. The fermionic zeromodes are used to
measure the topological charge $Q$ via the Atiyah-Singer index
theorem~\cite{Atiyah:1971rm,Schwarz:1977az,Brown:1977bj}
\begin{equation}
    \mathrm{ind}\;D[A] = n_- - n_+ = Q,
\end{equation}
where  $n_-$ and $n_+$ are the number of left- and right-handed zeromodes
of the Dirac operator $D$. This equation accounts for Wilson and overlap fermions in the fundamental representation. The adjoint version of the index theorem reads \begin{equation}
    \mathrm{ind}\;D[A] = n_- - n_+ = 2NQ = 4Q
\end{equation}
where $N=2$ is the number of colors and the additional factor $2$ is due to the fact that the fermion is in the real representation, hence the spectrum of the adjoint Dirac operator $iD$ is doubly degenerate. The eigenvalues of the staggered fermion operator have a twofold degeneracy due to a global charge conjugation symmetry in $SU(2)$. We therefore have $\mathrm{ind}\;D[A] = n_- - n_+ = 2Q$ for fundamental and $\mathrm{ind}\;D[A] = n_- - n_+ = 8Q$ for adjoint (asqtad) staggered fermions. 

\subsection{Fundamental zeromodes for intersecting center vortex fields with topological charge $Q=2$}\label{seq:fundzms}

\begin{figure}[htb]
\centering
\psfrag{x}{$x$}
\psfrag{y}{$z$}
\psfrag{z}{$z$}
a)\includegraphics[width=.44\linewidth]{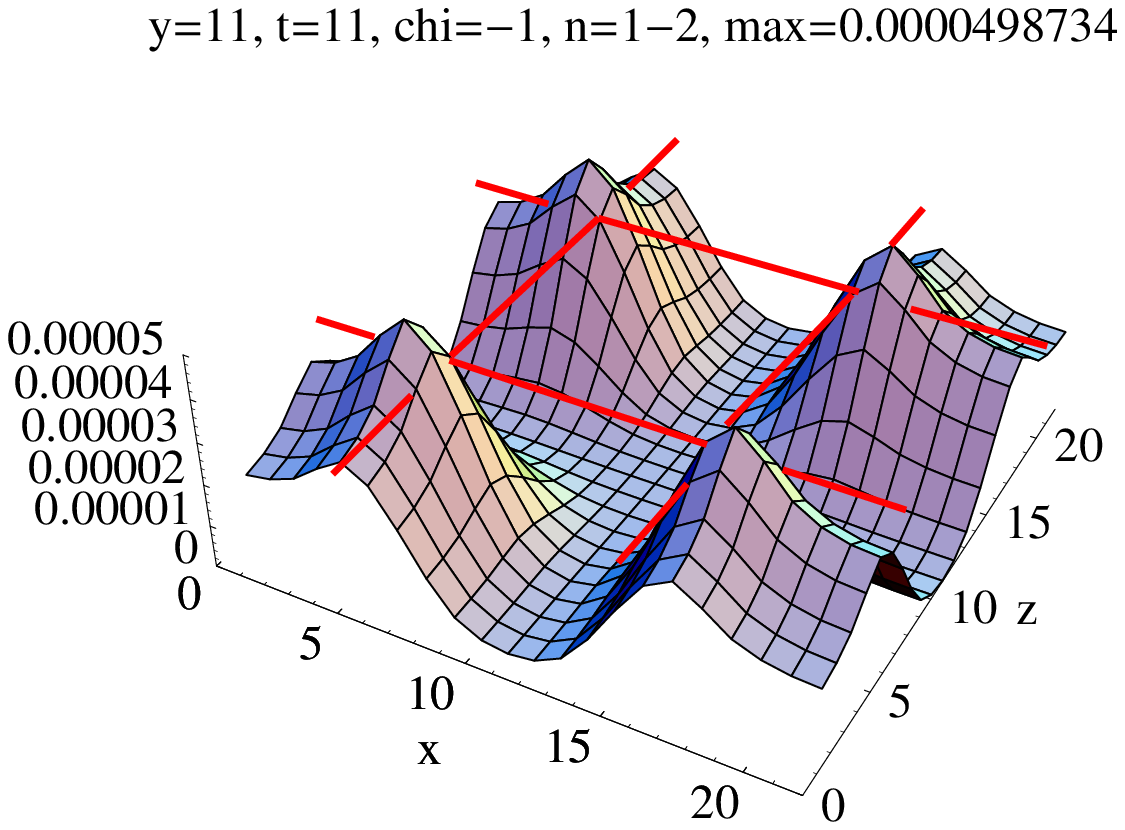}b)\includegraphics[width=.44\linewidth]{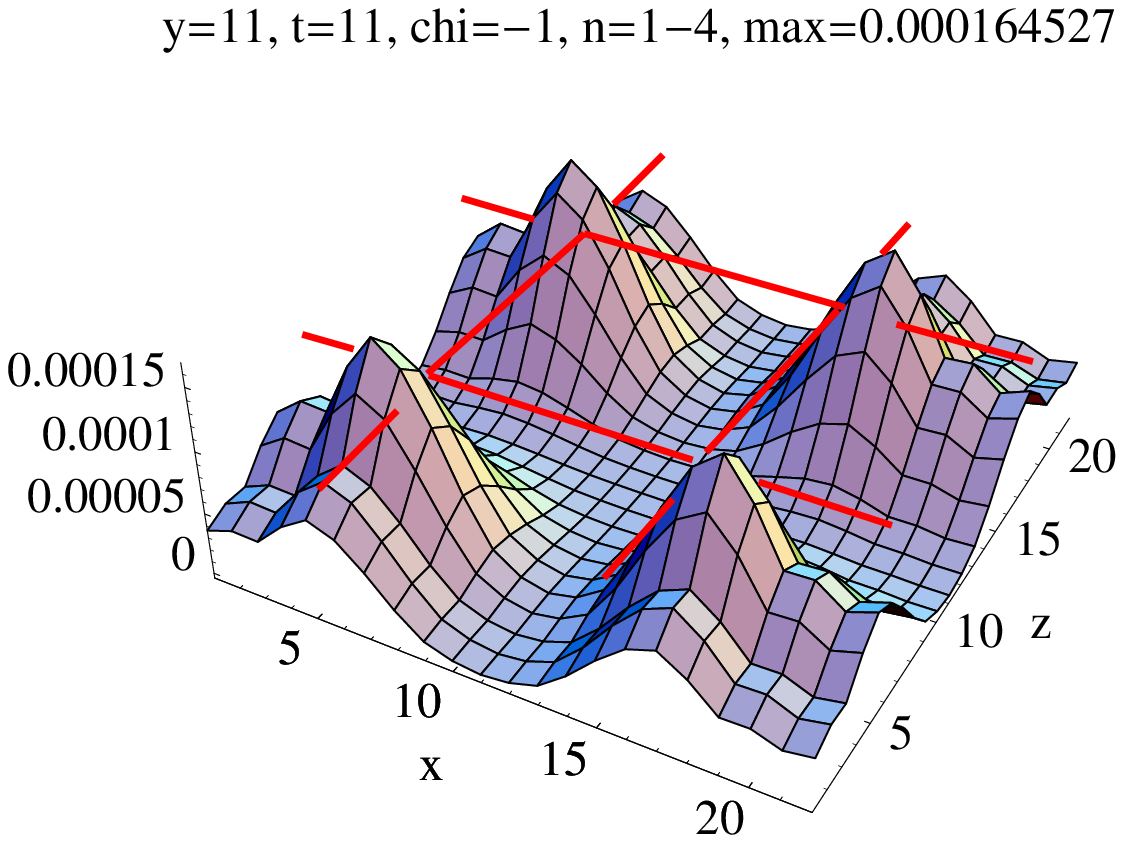}\\
c)\includegraphics[width=.44\linewidth]{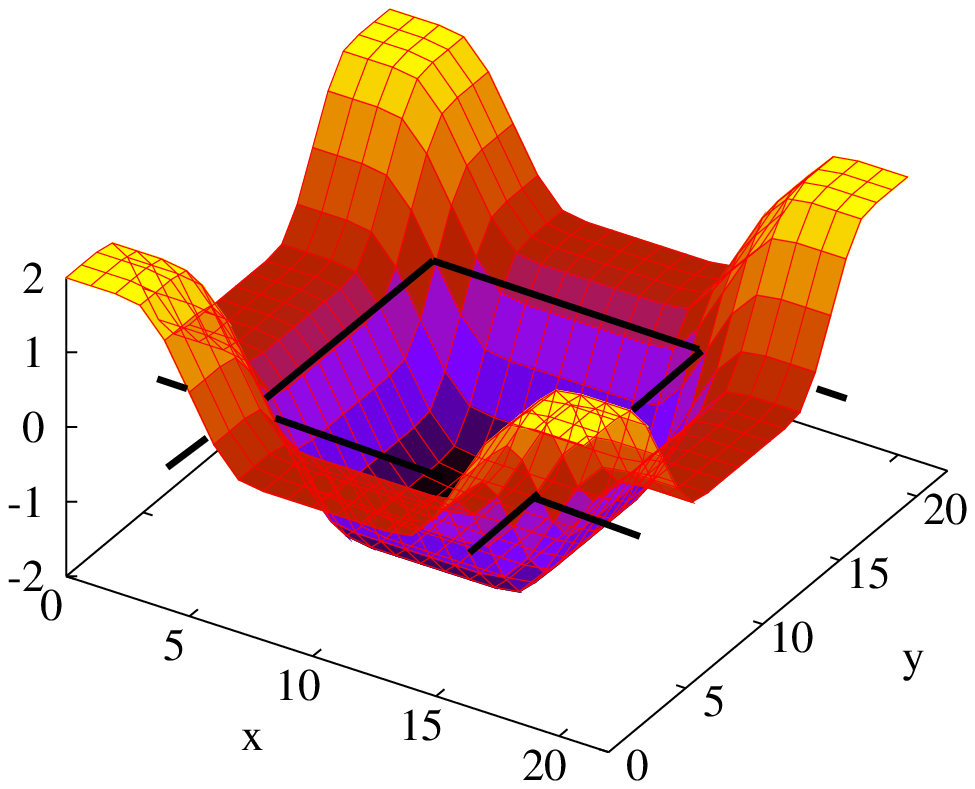}d)\includegraphics[width=.44\linewidth]{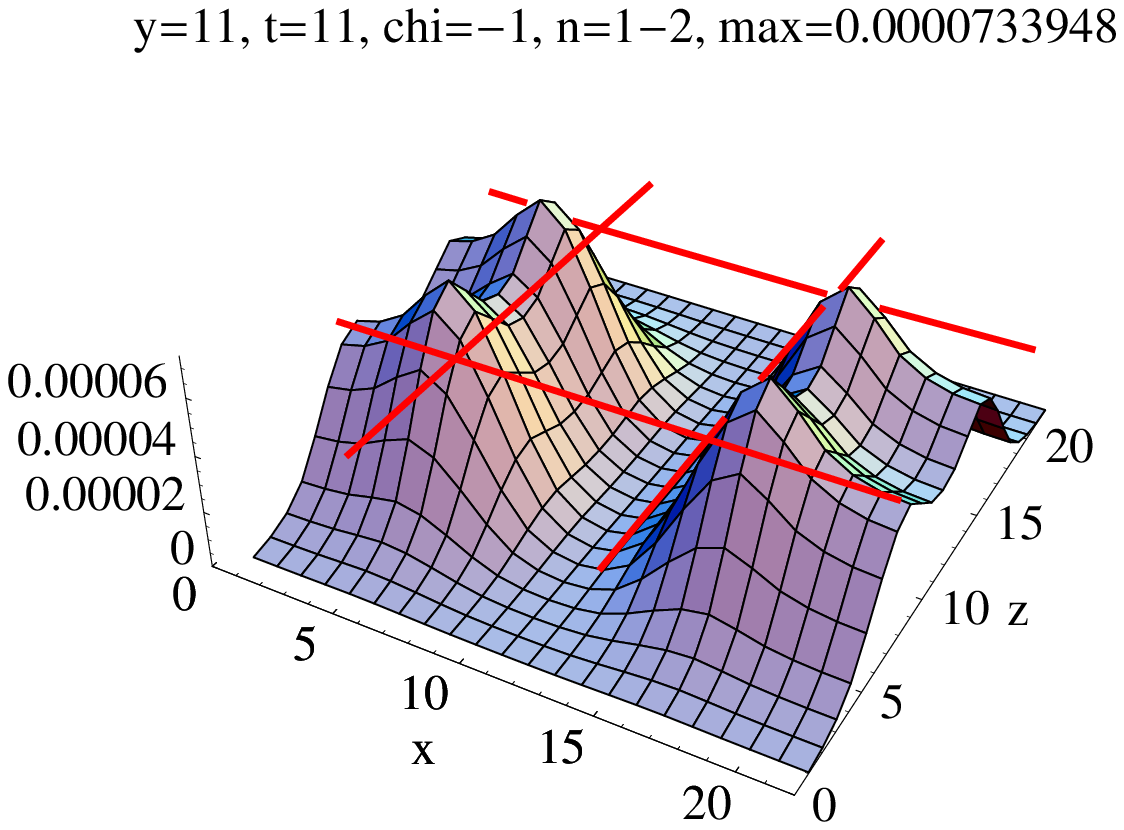}
\caption{$Q=2$ configuration: Scalar density plots of a) the two overlap zeromodes, b) the four asqtad staggered zeromodes, both with periodic boundary conditions, and d) the two overlap zeromodes with antiperiodic boundary conditions in time direction. The plot titles indicate the plane positions, the chirality (chi=$\pm1$) and numbers (n=1-2/n=1-4) of (overlap/asqtad staggered) zeromodes and the maximum density peak in the plot. c) Sum of Wilson lines in y- and t-direction (Polyakov-loops) in the intersection plane. P-vortices are indicated with red or black lines.}
\label{fig:plq2fund}
\end{figure}

We intersect two parallel vortex pairs with $x_1=z_1=6$ and $x_2=z_2=16$ at
$y=t=11$ respectively on a $22^4$-lattice. The four intersection points all
carry topological charge contributions of $+1/2$ and therefore sum up to a
total topological charge $Q=2$. In agreement with the lattice index theorem we
get two overlap and four asqtad staggered zeromodes of negative chirality
(left-handed) in the fundamental representation. Fig.~\ref{fig:plq2fund}
shows the scalar density plots of the fundamental overlap and asqtad
staggered zeromodes with periodic boundary conditions together with the sum of
Wilson lines in y- and t-direction (Polyakov-loops) in the intersection
plane as well as the scalar density plot of the two overlap zeromodes with
usual antiperiodic boundary conditions in time direction (asqtad staggered
modes again distribute similarly). The individual modes all distribute
equally, showing four distinct maxima, as trivially all their linear
combinations do. A close look shows that the zeromodes do not exactly peak at
the vortex intersections, they rather avoid regions with negative Polyakov
lines and approach the intersections (or the vortex surfaces) from regions
with positive Polyakov lines. This behavior was already observed
in~\cite{Jordan:2007ff} for spherical vortices, the zeromodes avoid regions
with negative Polyakov lines. The sensitivity of Dirac eigenmodes to the
Polyakov value can be exhibited clearly using one parallel vortex pair (e.g. two parallel $xy$-vortices), which are exchanged by a discrete translation $T_z$ in the $z$-direction by half the lattice length $N_z$. The configuration apparently consists of two identical flux lines because the field strength is invariant under the $z$-translation $T_z$. However, this operation is not a symmetry of the Dirac operator since it would change the Polyakov loop, which is a gauge-invariant quantity. Therefore the distribution of the scalar density of the Dirac eigenfunctions is not invariant under $T_z$. In fact, the antiperiodic boundary conditions in time direction change the time-like Polyakov lines and the zeromode density shows the shift in the $z$-direction by half the lattice size, see Fig.~\ref{fig:plq2fund}d. 

The configuration is also presumably equivalent to the abelian gauge fields whose zeromodes are derived analytically in~\cite{Reinhardt:2002cm}. Nevertheless, the probability distribution in the intersection plane shown in the reference differs significantly from the one we obtained, namely in that the former exhibits exactly the translation symmetry discussed above. This may indicate a relation to the issue of the Polyakov loop because configurations having the same flux distribution may still differ in the values of their Polyakov loops. The values of the Polyakov loops in our study and the calculation of \cite{Reinhardt:2002cm} only coincide in the intersection plane. The choice of boundary conditions is another possibility for the origin of the discrepancy. For the flux of two parallel center vortices the transition matrices need to have a winding number one around the boundary. On the lattice, the transition functions are supposed to be incorporated in the links such that all plaquettes agree with the corresponding integrals of the continuum field. Hence it should be sufficient to use periodic boundary conditions for the links as well as for the fermions. However, this allows further zeromodes which cancel each other in the index theorem (non-topological zeromodes) and so we prefer the usual antiperiodic boundaries in time direction which only reproduce the relevant zeromodes.

\subsection{Adjoint zeromodes for intersecting center vortex fields with topological charge $Q=0$ and $Q=2$}

Now we try to locate the fractional topological charge contributions with adjoint eigenmodes of the overlap Dirac operator which are also sensitive to topological charge contributions of $|Q|=1/2$. For asqtad staggered fermions we find the correct (doubled) numbers of zeromodes in all cases and the scalar densities of the sum of all zeromodes always localizes similarly. 

We do not find zeromodes localized to a single region with nonvanishing topological charge contributions. Therefore we use the inverse participation ratio (IPR)~\cite{Ilgenfritz:2007xu, Polikarpov:2005ey, Aubin:2004mp, Bernard:2005nv} to quantify the localization of eigenmodes. The IPR of a normalized ($\sum_x\rho_i(x)=1$) field $\rho_i(x)$ is defined as
\begin{equation}
I=N\sum_{x=0}^N\rho_i^2(x)
\end{equation}
where N is the number of lattice sites $x$. With this definition, $I$ characterizes the inverse fraction of sites contributing significantly to the support of $\rho(x)$, {\it i.e.}, a high IPR indicates that the eigenmode is localized to a a few lattice points only. We perform systematic and random IPR maximization procedures for linear combinations of zeromodes in order to get single eigenmode peaks localized to regions with nonvanishing topological charge contribution.


{\bf Q=0:}

\begin{figure}[htb]
\centering
\psfrag{x}{$x$}
\psfrag{z}{$z$}
\psfrag{-6}{}
\psfrag{10a}{\scriptsize $10^{-6}$}
a)\includegraphics[width=.3\linewidth]{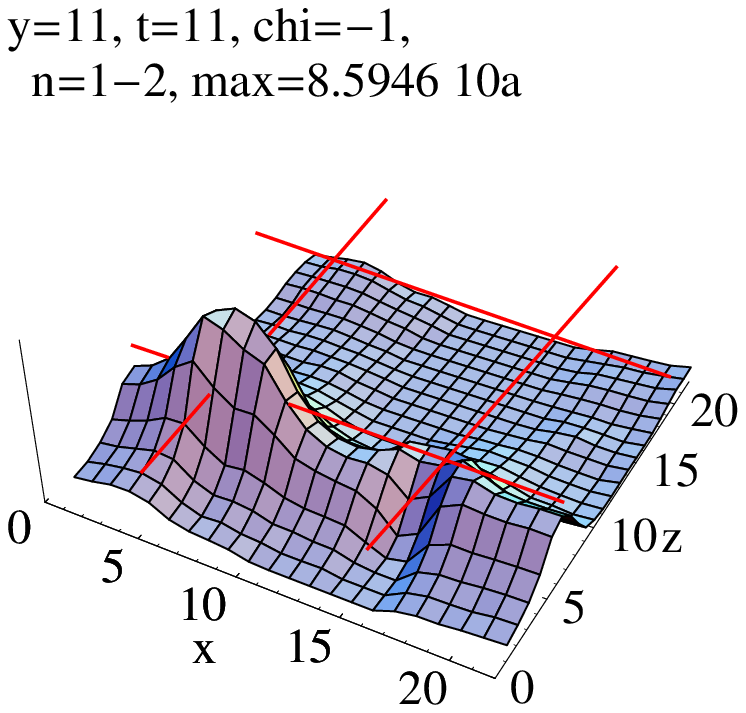}b)\includegraphics[width=.3\linewidth]{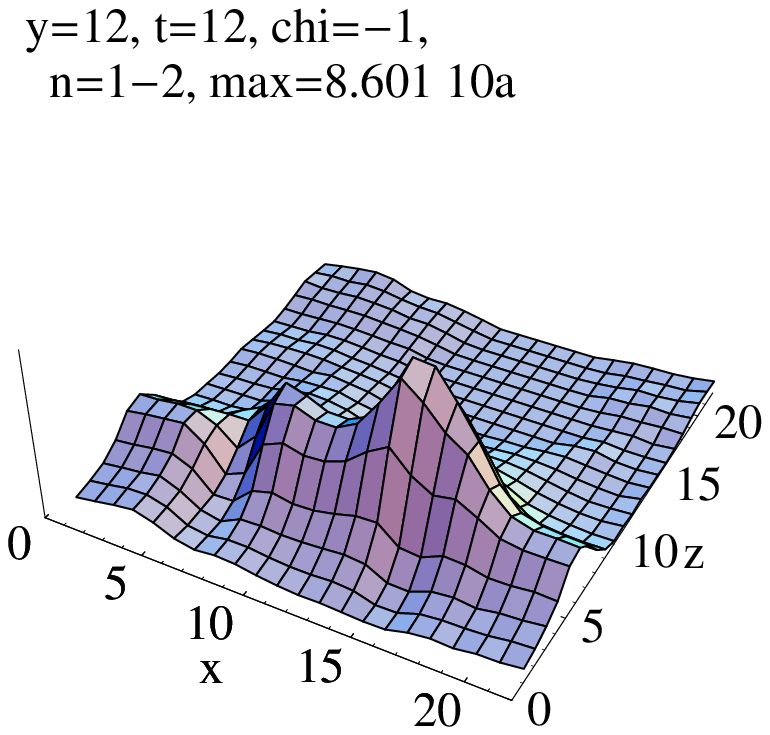}c)\includegraphics[width=.3\linewidth]{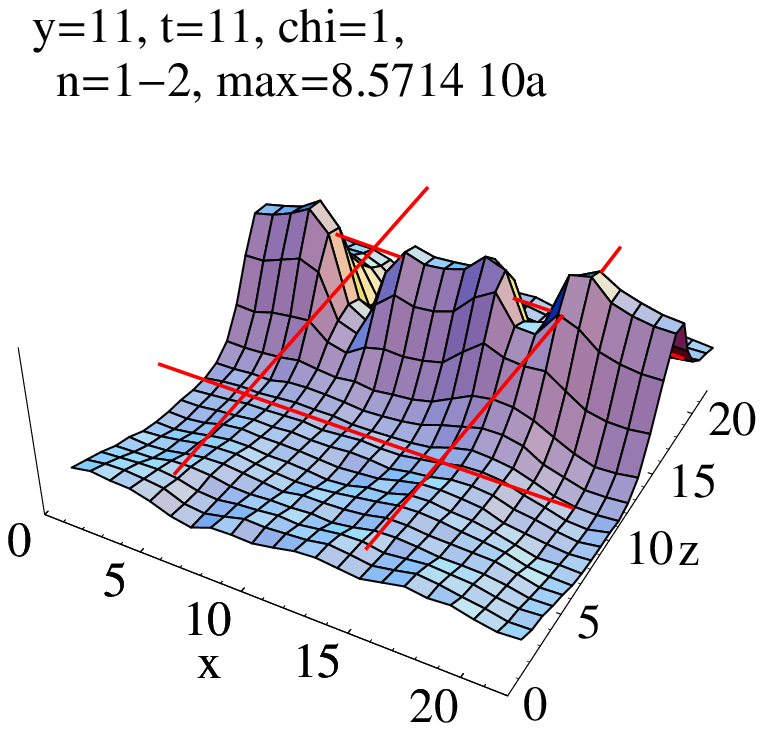}
\caption{$Q=0$ configuration: Scalar density plots of the two adjoint overlap zeromodes (n=1-2) with negative chirality (chi=-1) in the $xy$-plane at a) $y=t=11$ (intersection plane, red lines indicate the P-vortices), b) $y=t=12$ and c) the two right-handed (chi=1) adjoint overlap zeromodes at $y=t=11$.}
\label{fig:plq0adj}
\end{figure}

For the configuration with topological charge $Q=0$ we intersect two anti-parallel vortices with the same vortex centers as for the $Q=2$ configuration described above ($x_1=z_1=6$ resp. $x_2=z_2=16$) at $y=t=11$ respectively on a $22^4$-lattice. For this configuration we do not get fundamental zeromodes, but with periodic boundary conditions we find two adjoint overlap (four asqtad staggered) zeromodes of each chirality. Fig.~\ref{fig:plq0adj} shows the scalar density plots of the adjoint overlap zeromodes for the $Q=0$ configuration. The left-handed zeromodes (Fig.~\ref{fig:plq0adj}a) peak at the intersection point $(6,11,6,11)$ of topological charge $Q=1/2$ with a maximum density of $8.595\cdot10^{-6}$ but a second maximum can be found near the intersection plane at $(16,12,6,12)$ with a maximum value of $8.601\cdot10^{-6}$ (Fig.~\ref{fig:plq0adj}b) which is next to the intersection with $Q=-1/2$ at $(16,11,6,11)$. These zeromodes rather localize the $xy$-vortex sheet at $z=6$, pronouncing the intersections whereas the right-handed zeromodes of Fig.~\ref{fig:plq0adj}c locate the other $xy$-vortex sheet at $z=16$, but with notches at the two intersection points. The individual zeromodes of the same chirality show identical scalar density distributions, therefore it is not possible to find linear combinations of the zeromodes which locate single vortex intersections.

{\bf Q=2:}

\begin{figure}[p]
\centering
\psfrag{x}{$x$}
\psfrag{z}{$z$}
a)\includegraphics[width=.44\linewidth]{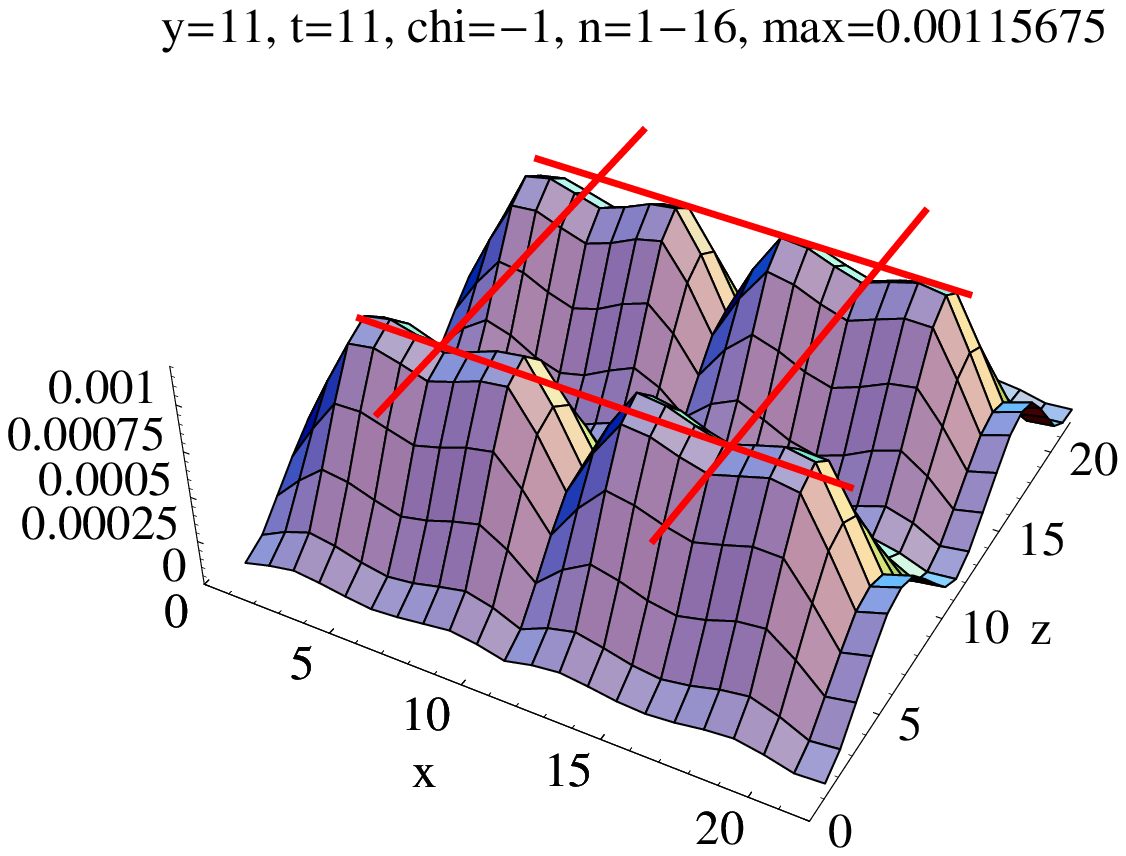}b)\includegraphics[width=.44\linewidth]{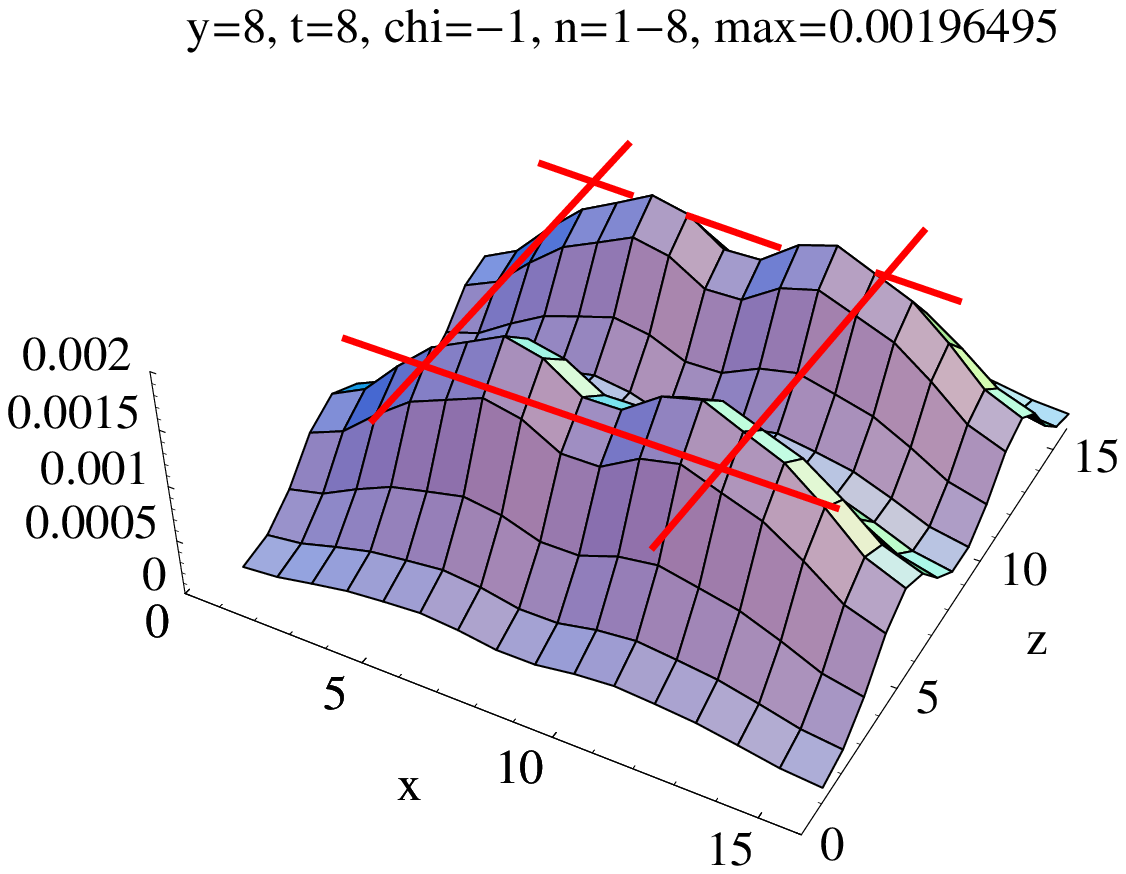}
\textcolor{white}{emty line}\\
c)\includegraphics[width=.44\linewidth]{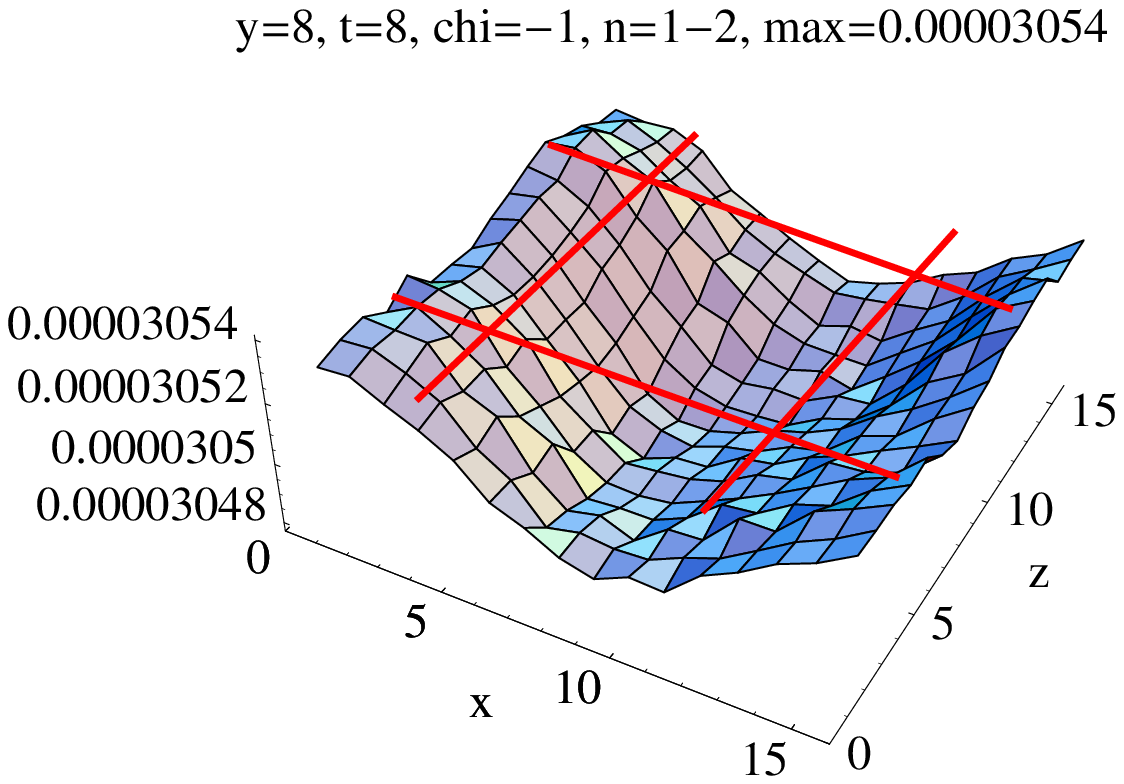}d)\includegraphics[width=.44\linewidth]{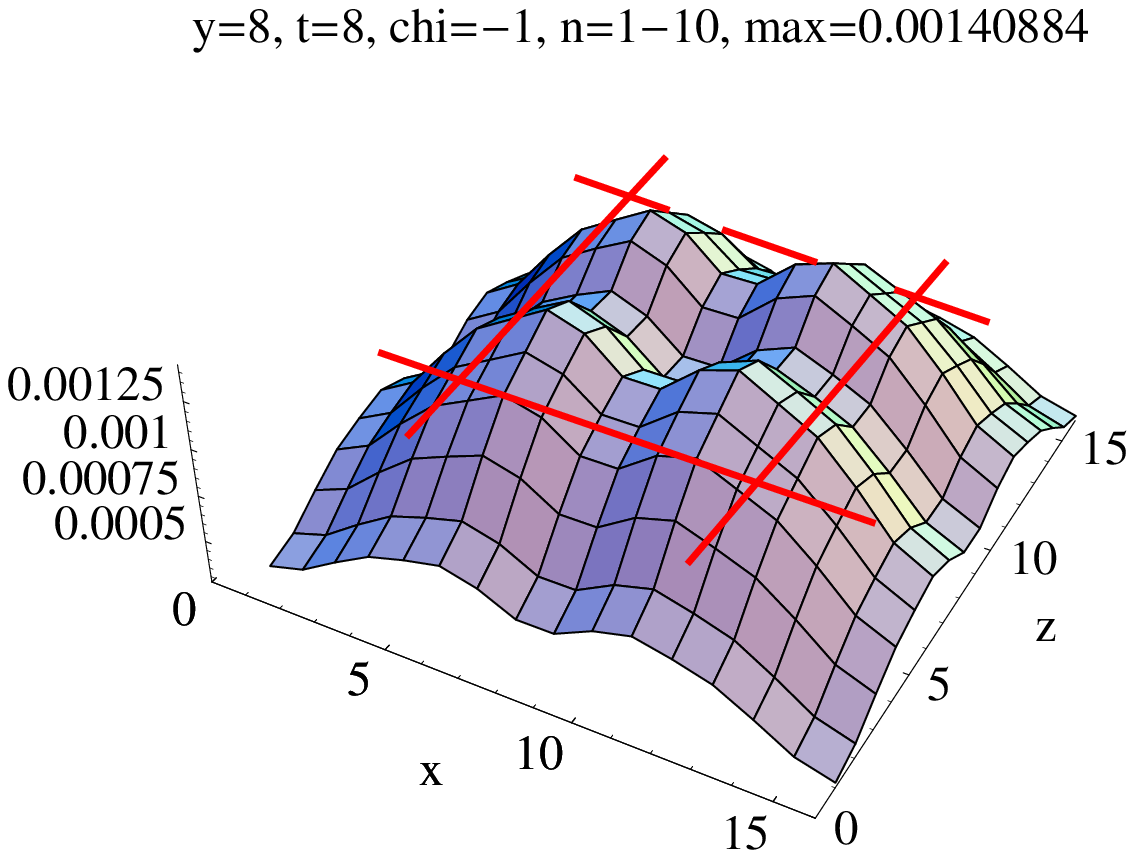}
\psfrag{IPR}[r][t][.8][0]{IPR}
e)\includegraphics[width=.52\linewidth]{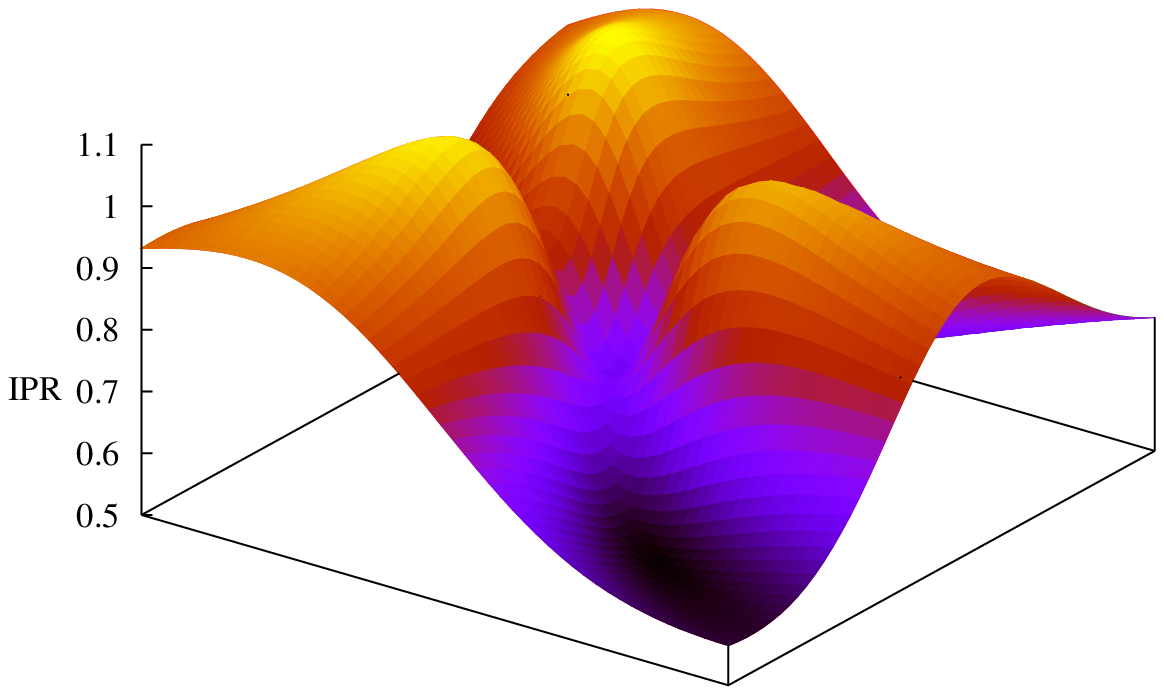}f)\includegraphics[width=.48\linewidth]{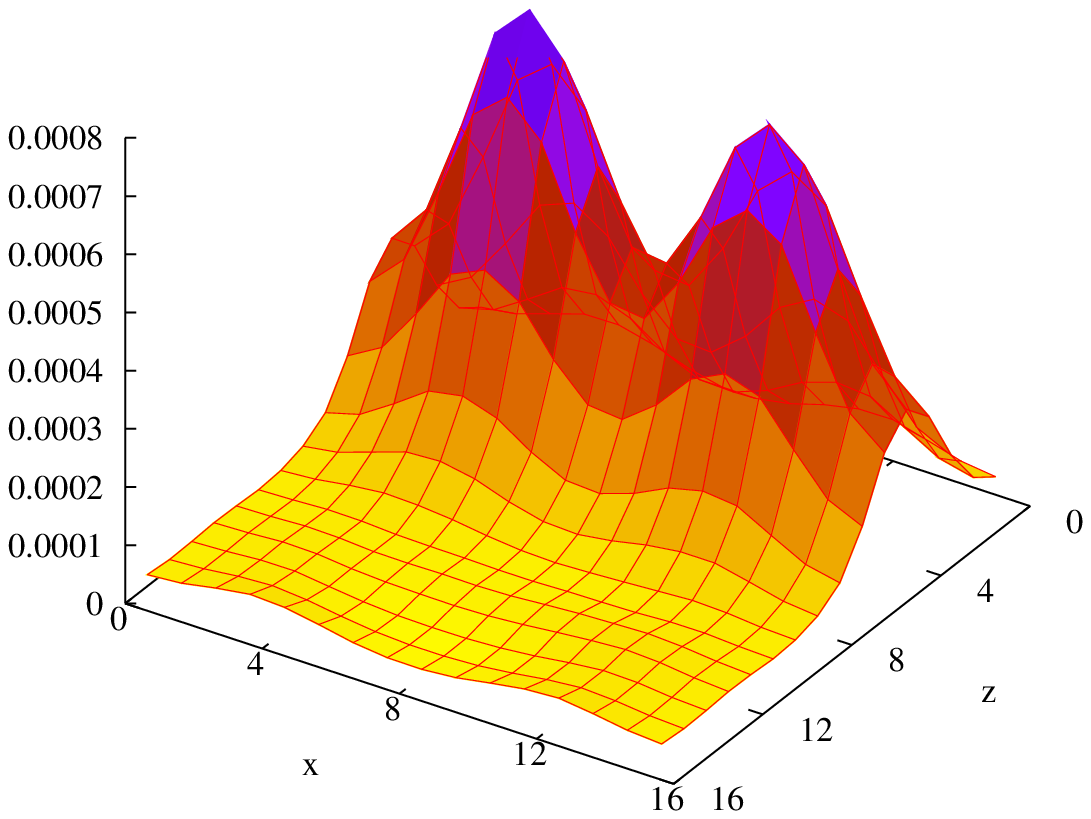}
\caption{$Q=2$ configuration: Scalar density plots of the a) 16 asqtad staggered and b) eight overlap left-handed adjoint zeromodes for antiperiodic boundary conditions, the modes locate the $xy$-vortex pair. Further, c) two right-handed (non-topological) and d) ten left-handed adjoint overlap zeromodes for periodic boundary conditions. e) Plane through the highest three IPR maxima in the parameter space of linear combinations of the eight zeromodes. The peaks are very broad and easy to identify. f) The scalar density of the linear combination of the eight zeromodes with maximal IPR still peaks at two vortex intersections.}
\label{fig:plq2adj}
\end{figure}

The configuration with topological charge $Q=2$ on the $22^4$-lattice from above gives 16 adjoint asqtad staggered zeromodes with negative chirality for antiperiodic boundary conditions in time direction. For adjoint overlap fermions we use a $16^4$-lattice with vortex centers $x_1=z_1=4$ and $x_2=z_2=12$ at $y=t=8$ respectively and we find eight left-handed zeromodes. With periodic boundary conditions we find two positive (non-topological) and ten negative zeromodes. Fig.~\ref{fig:plq2adj} shows the scalar density plots of adjoint overlap and asqtad staggered zeromodes of the $Q=2$ configuration. The zeromodes locate the $xy$-vortex pairs but even individual modes do not show single peaks locating one intersection with $Q=1/2$. Linear combinations of the eight zeromodes with negative chirality show six distinct IPR maxima. This was obtained by a systematic study with the eight coefficients of the linear combination varying from $-10$ to $10$ in integer steps. Further we started from 20.000 random points in the parameter space and determined the nearest maximum by the gradient method. Each of the six maxima was found between $2.000-6.000$ times and no other maxima were obtained. In Fig.~\ref{fig:plq2adj}e we plot a 2D cut through the first three IPR maxima in the 8D parameter space of linear combinations. The scalar density of the linear combination of the eight zeromodes with maximal IPR is presented in Fig.~\ref{fig:plq2adj}f, it still peaks at two vortex intersections. The ten zeromodes for periodic boundary conditions look more symmetric, but again no single mode locates one vortex intersection. We again performed a systematic search with the ten coefficients of the zeromodes varying from $-6$ to $6$ in integer steps and again $20.000$ random points. In this case we found only five maxima, each $3.000-5.000$ times. Again, it is not possible to find linear combinations of the zeromodes to locate at single vortex intersections.

\subsection{Adjoint zeromodes for center vortex fields with one single "thick" intersection and topological charge $Q=1/2$}

\begin{figure}[htb]
\centering
\psfrag{x}{$x$}
\psfrag{z}{$z$}
\psfrag{P}[c][c][.8][0]{$P$}
\psfrag{profile}[r][c]{\scriptsize $\phi(x)$}
\psfrag{polline}[r][c]{\scriptsize tr$U_y(x)$}
\psfrag{sigma3}[r][c]{\scriptsize $\sigma_3$-comp.}
\psfrag{2d}{\scriptsize $d$}
\psfrag{1}{\scriptsize $1$}
\psfrag{0}{\scriptsize $0$}
\psfrag{3}{\scriptsize $3$}
\psfrag{f}{\textcolor{blue}{$\phi$}}
\psfrag{12}{\scriptsize $N_z$}
\psfrag{p}{$\pi$}
\psfrag{-1}{\scriptsize $-1$}
\psfrag{t}{\scriptsize \textcolor{red}{${\bf Tr} U_A$}}
a)\includegraphics[width=.44\linewidth]{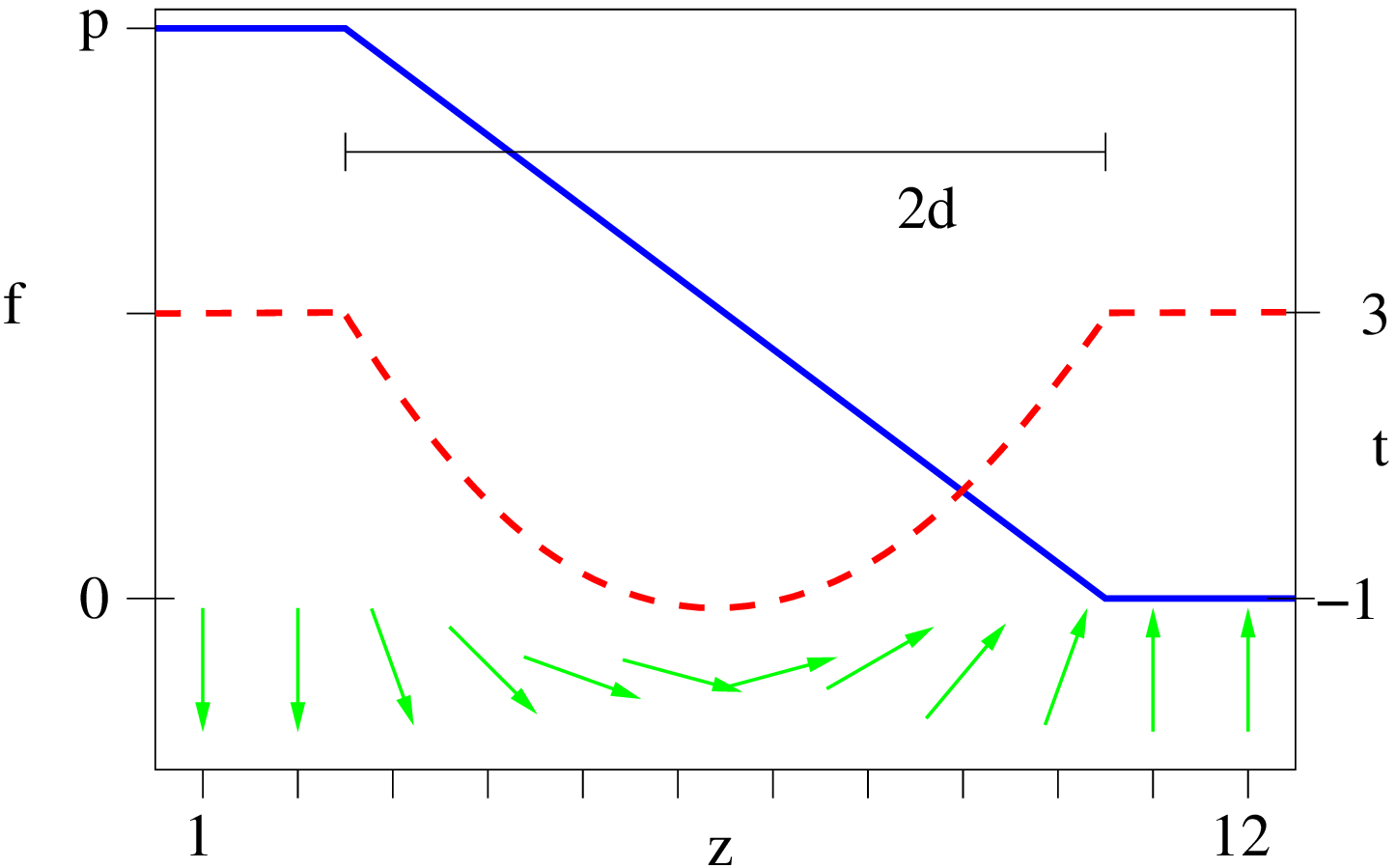}$\qquad$ b)\includegraphics[width=.4\linewidth]{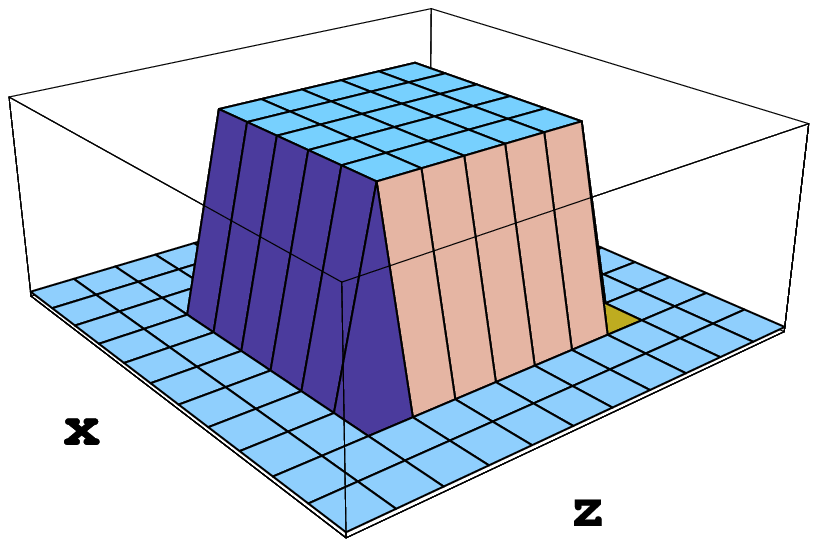}
\caption{"$Q=1/2$ configuration": a) Link profile of a "thin-thick" plane vortex, the link angle $\phi$ (blue) decreases from $\pi$ to $0$ within a certain vortex thickness $d$. The arrows (links) rotate counterclockwise with decreasing $\phi$. The thin vortex is given by the jump at the boundary. The red dashed line shows the trace of adjoint links ${\bf Tr} U_A$ (see text below) b) Topological charge density in the intersection plane.}
\label{fig:1hlfconfig}
\end{figure}

We analyze a configuration of "thin-thick" vortex intersections apparently
having topological charge $|Q|=1/2$. The profile of a "thin-thick" vortex is
plotted in Fig.~\ref{fig:1hlfconfig}a. The thin vortex sheet is defined by the jump of the $y$- or $t$-link from $+1$ to $-1$ at the boundary. The thick
vortex is located symmetrically around the center of the lattice with thickness $d$. The thin-thick $xy$- and $zt$-vortices still intersect at four points, but the plaquette or hypercube definitions of topological charge do not recognize the thin vortex sheets and therefore only measure one topological charge contribution $Q=1/2$ of the "thick" vortex intersection, see Fig.~\ref{fig:1hlfconfig}b. 

\begin{figure}[p]
\psfrag{x}{$x$}
\psfrag{z}{$z$}
\psfrag{y}{$y$}
\psfrag{t}{$t$}
\psfrag{10a}{\scriptsize $10^{-7}$}
\centering
a)\includegraphics[width=.44\linewidth]{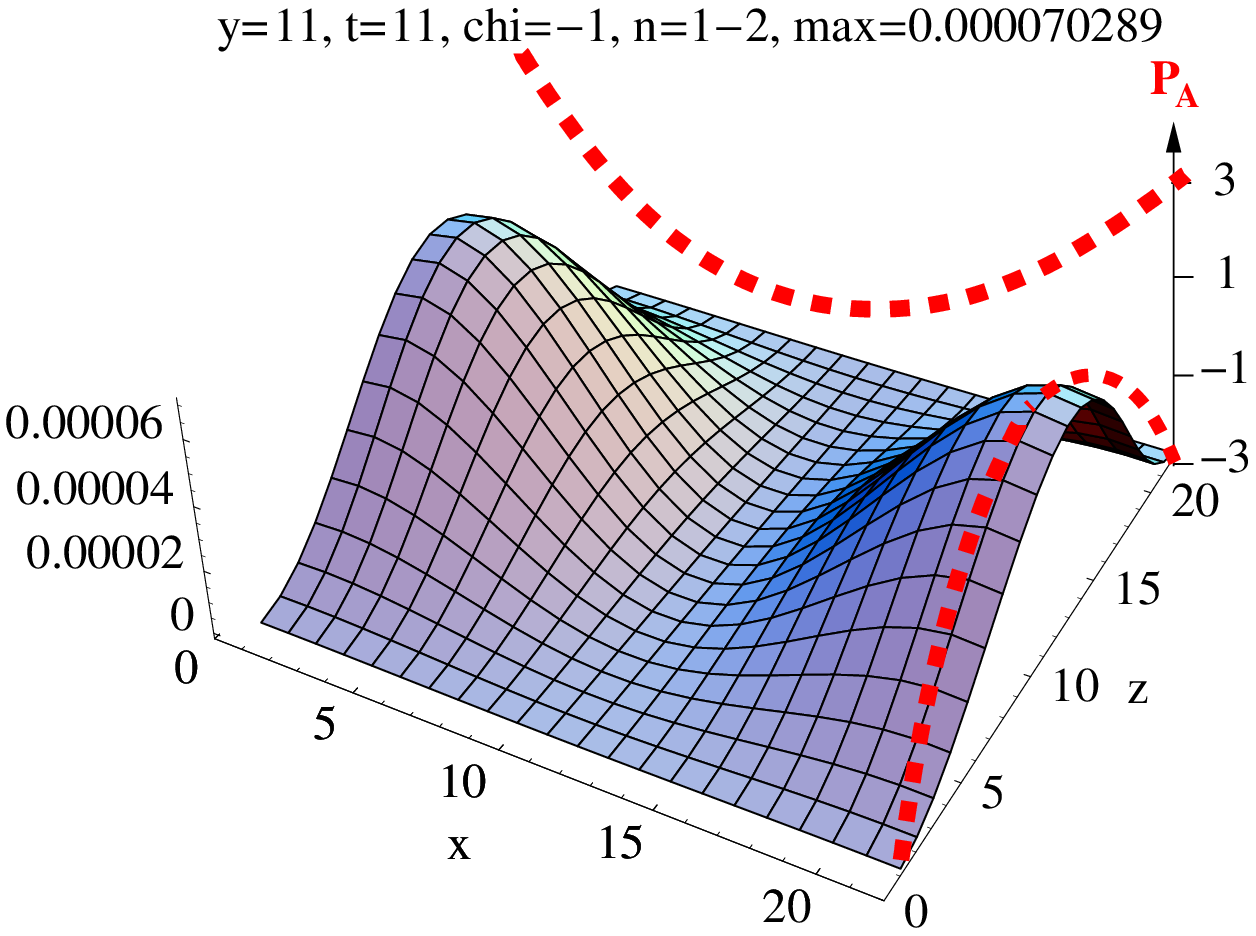}b)\includegraphics[width=.44\linewidth]{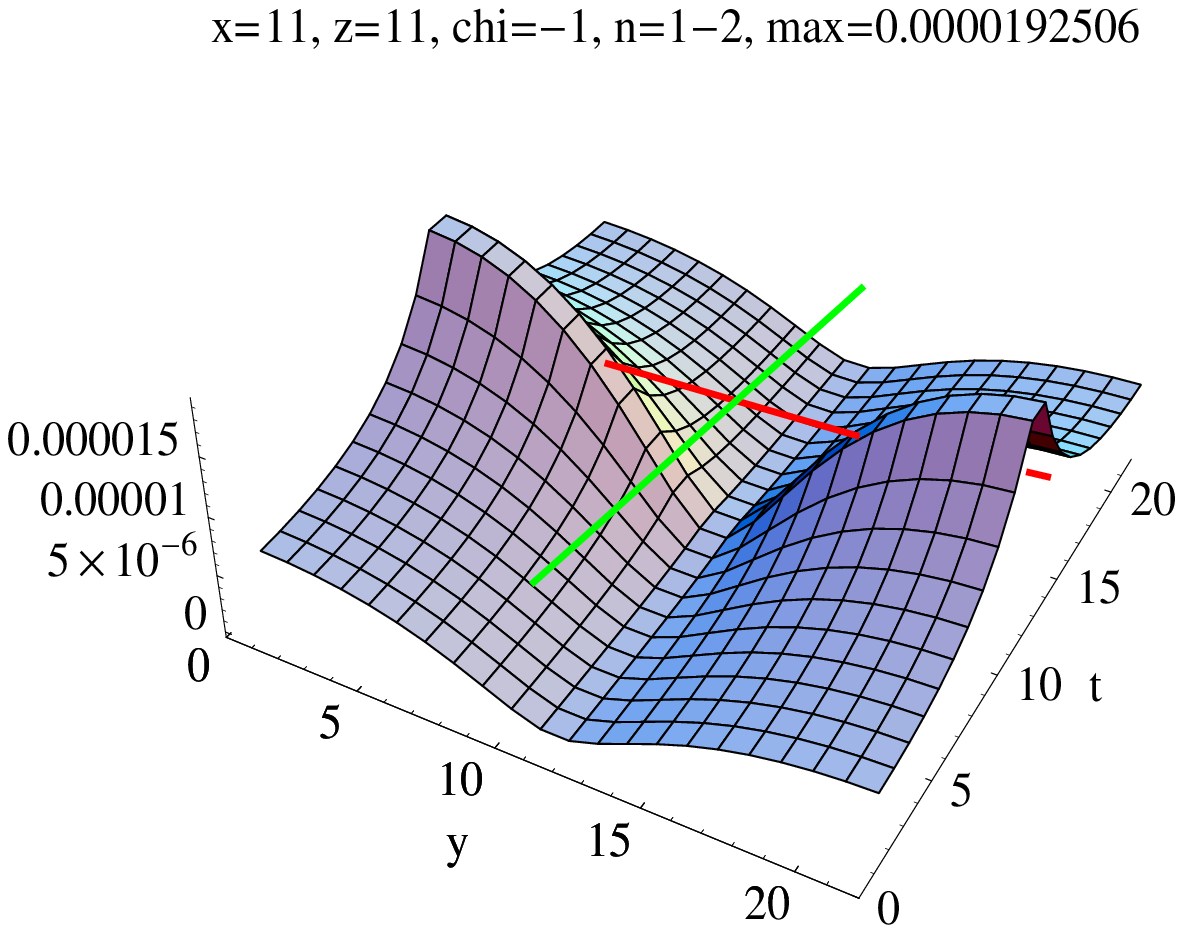}\\
\textcolor{white}{emty line}\\
c)\includegraphics[width=.435\linewidth]{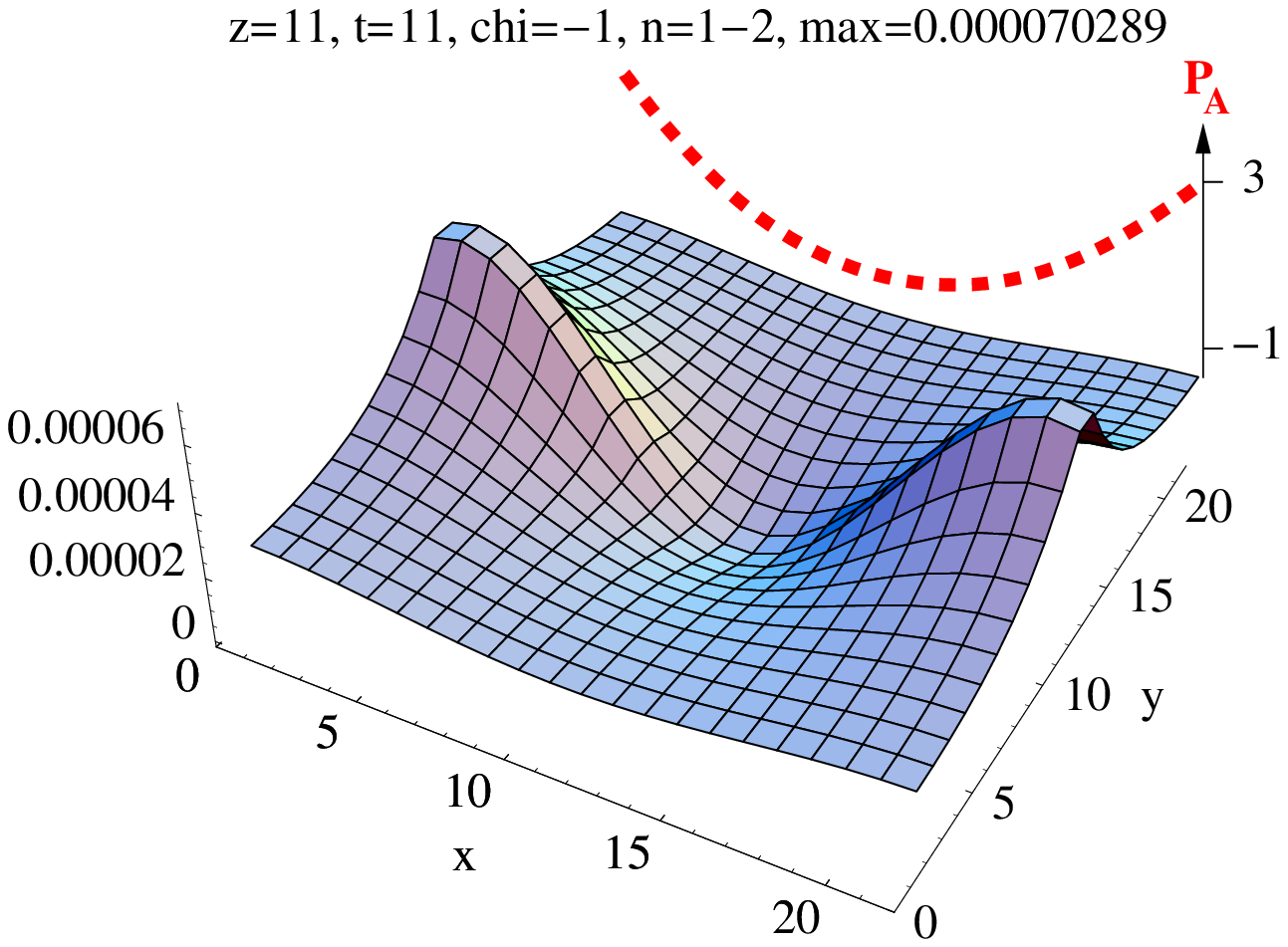}d)\includegraphics[width=.44\linewidth]{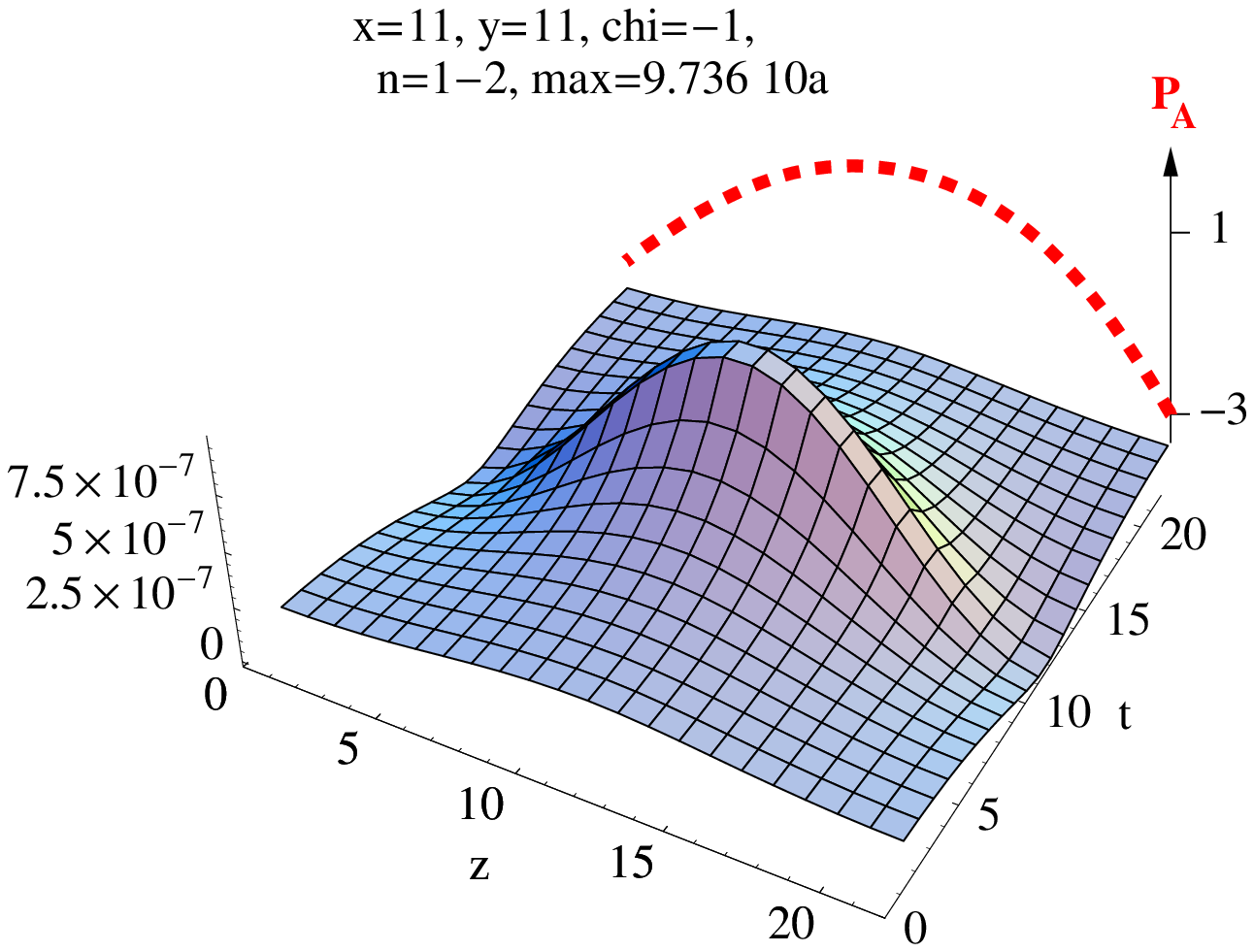}\\
\textcolor{white}{emty line}\\
\caption{"$Q=1/2$ configuration": Scalar eigenmode density of two adjoint overlap (identical to four adjoint asqtad staggered) zeromodes of negative chirality in various planes through the thick-thick vortex intersection: a) $xz$-plane (intersection plane): The zeromodes avoid regions of negative adjoint Polyakov (Wilson) lines ($P_A$, red dots) with respect to boundary conditions (see text above, Eq.~(\ref{eq:adjtrace}) and Fig.~\ref{fig:1hlfconfig}a) and therefore do not peak at the topological charge contribution $Q=1/2$ of the "thick-thick" vortex intersection at $x=z=11$; the "thin-thin" and "thin-thick" intersections are not recognized by the Dirac modes; b) $yt$-plane: The antiperiodic boundary conditions invert the profile in time- compared to the one in $y$-direction, hence the zeromodes prefer the $xy$-vortex at $t=11$ and avoid the $zt$-vortex at $y=11$ (the lines indicate the $xy$ (red) and $zt$ (green) vortex sheets in the $t=11$ resp. $y=11$ slices); c) $xy$-plane: The zeromodes reflect the profile of the adjoint $y$-Wilson lines ($P_A$, red dots) of the $zt$-vortex in $x$-direction at $y=11$; d) $zt$-plane: The zeromodes reflect the profile of the adjoint Polyakov lines ($P_A$, red dots) of the $xy$-vortex in $z$-direction at $t=11$, inverted by the antiperiodic boundary conditions in time direction.}
\label{fig:1hlfmodes}
\end{figure}

We compute the Dirac eigenmodes on a $22^4$-lattice with two intersecting
"thin-thick" $xy$- and $zt$-vortices with vortex thickness $d=20$ at
$t=y=11$, respectively. We get no fundamental zeromodes for the above topological
charge "$|Q|=1/2$ configuration". But the adjoint Dirac operator does not recognize the thin
vortices, just like the field theoretic operators we used to measure topological charge density. We find two adjoint overlap and four adjoint staggered zeromodes with negative chirality, which due to the index theorem again results in $Q=1/2$. For the adjoint fermions this configuration truncates three of the four vortex intersections and in this way simulates a 
situation related to the one achieved by twisted boundary conditions,
namely, a single detectable intersection. Therefore it is possible to have a
configuration that looks like having fractional topological charge. The eigenmode density distributions of overlap and asqtad staggered zeromodes are identical, Fig.~\ref{fig:1hlfmodes} shows the former for the intersection plane and for orthogonal planes to it at the intersection point. The modes are clearly sensitive to the traces of the adjoint link $U_A$ (see also Fig.~\ref{fig:1hlfconfig}a)
\begin{equation}
({\bf Tr} U)({\bf Tr} U)^\dagger=({\bf Tr} U)^2=1+{\bf Tr} U_A,
\label{eq:adjtrace}
\end{equation}
which in our case define the adjoint Polyakov lines $P_A$ (Wilson lines), since we have only nontrivial $t$-links in one time-slice and nontrivial $y$-links in one $y$-slice. The zeromodes prefer regions of positive Polyakov lines and avoid negative Polyakov lines with respect to the boundary conditions. For the antiperiodic boundary conditions in time direction the signs of the Polyakov lines are exchanged. Hence, the zeromode densities peak at the $xy$-vortex center at $z=t=11$ and avoid the $zt$-vortex center at $x=y=11$, or rather peak at the boundary parallel to the $zt$-vortex at $x=0$ or $x=22$ in the $y=11$-plane. 
For completeness we should mention that we get exactly the same results if we
analyze a "$Q=-1/2$ configuration", which can be realized if we rotate the links of one of the thick vortices in the opposite direction. In this case we get two adjoint overlap and four adjoint staggered zeromodes of {\it positive} chirality which show the same behavior.

\begin{figure}[p]
\centering
\psfrag{x}{$x$}
\psfrag{z}{$z$}
a)\includegraphics[width=.44\linewidth]{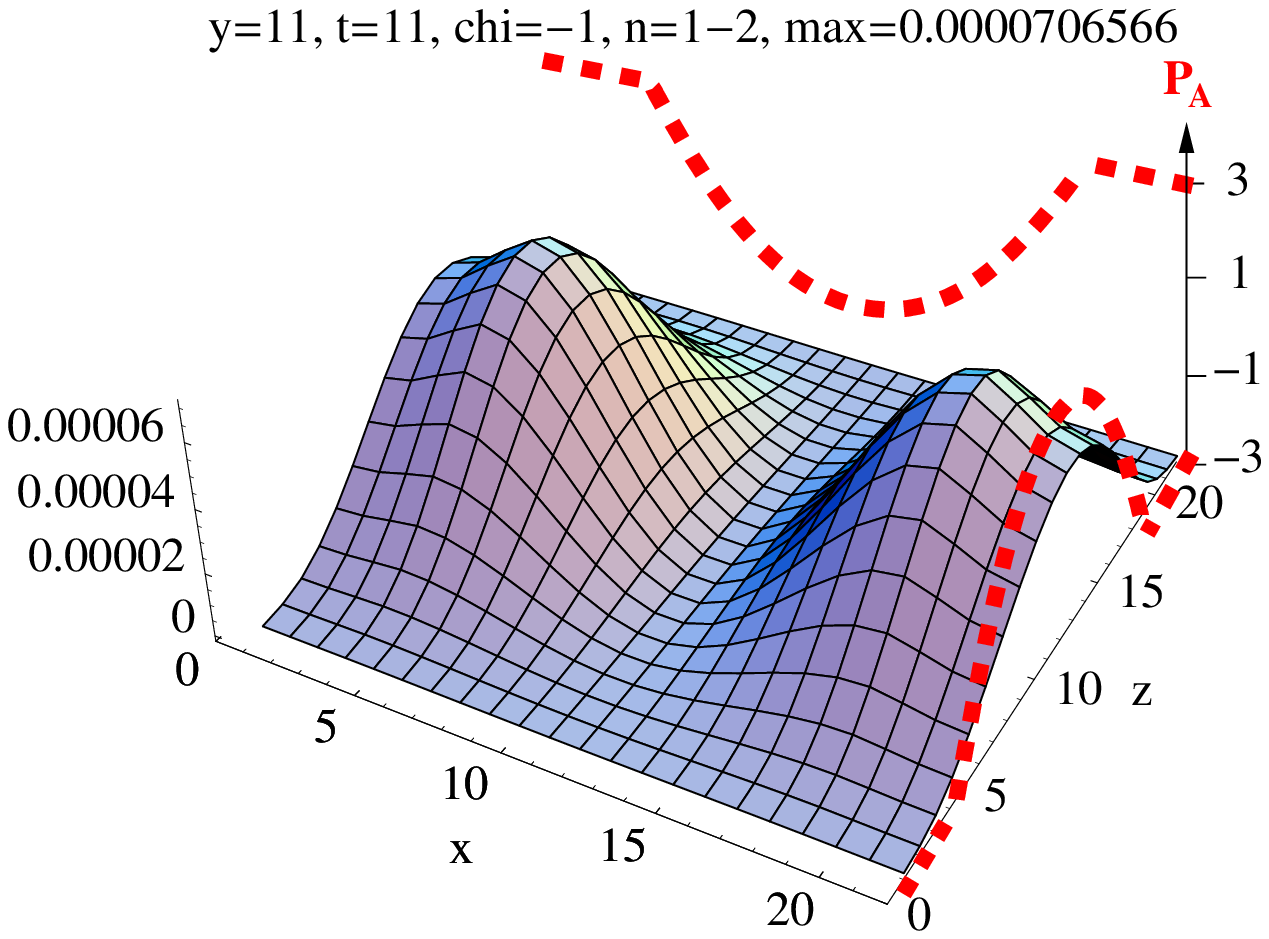}b)\includegraphics[width=.44\linewidth]{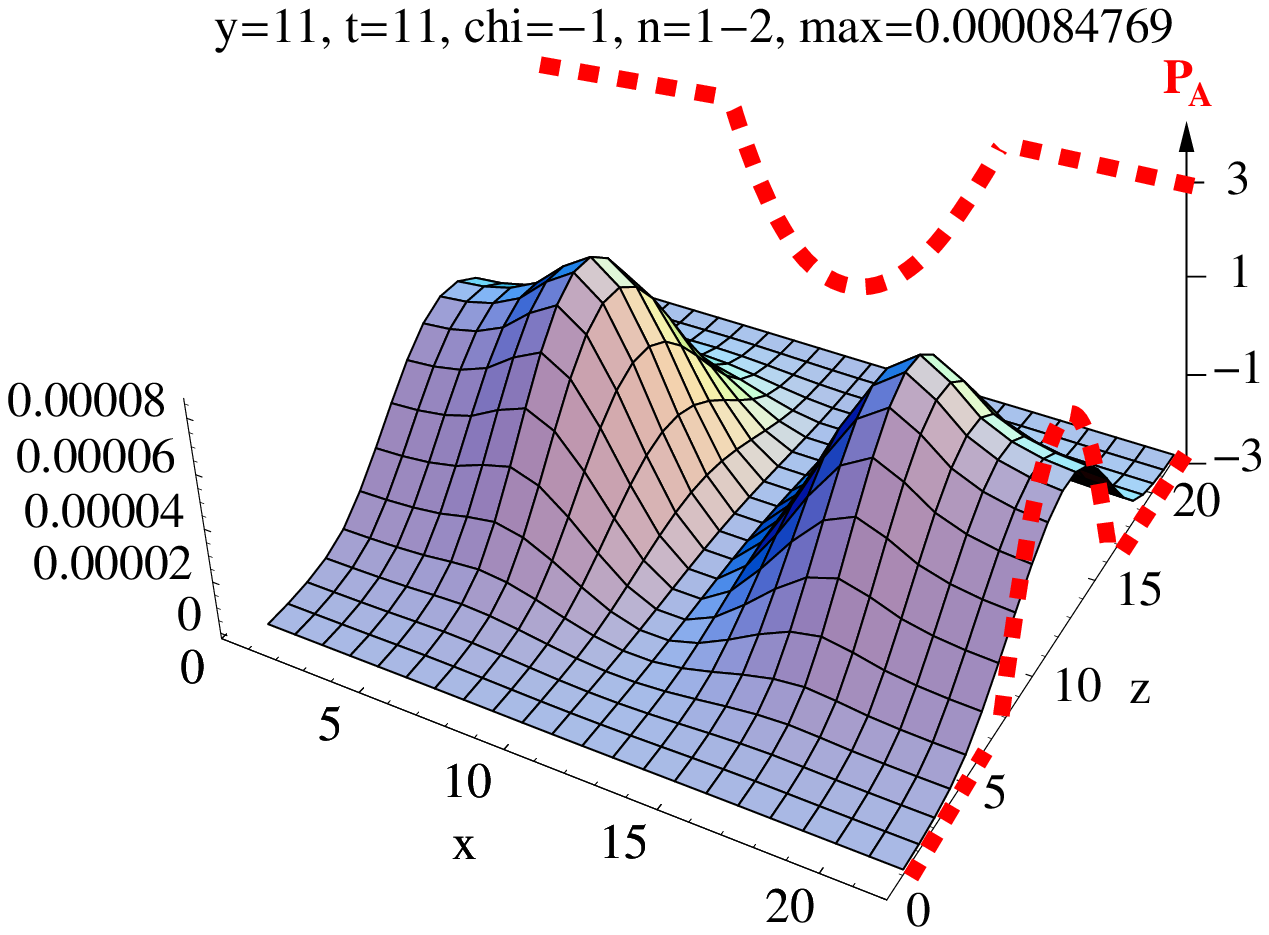}
\textcolor{white}{emty line}\\
c)\includegraphics[width=.44\linewidth]{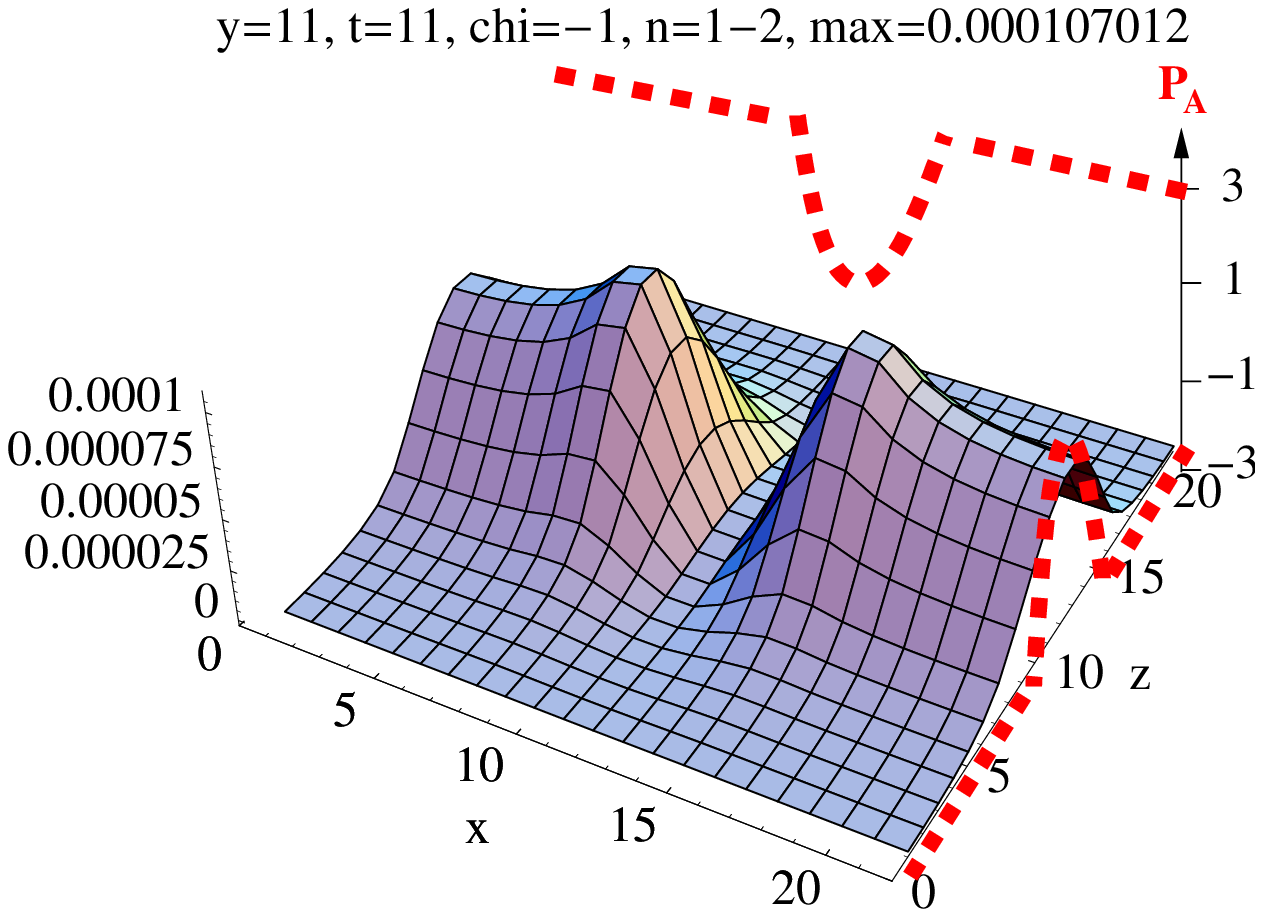}d)\includegraphics[width=.44\linewidth]{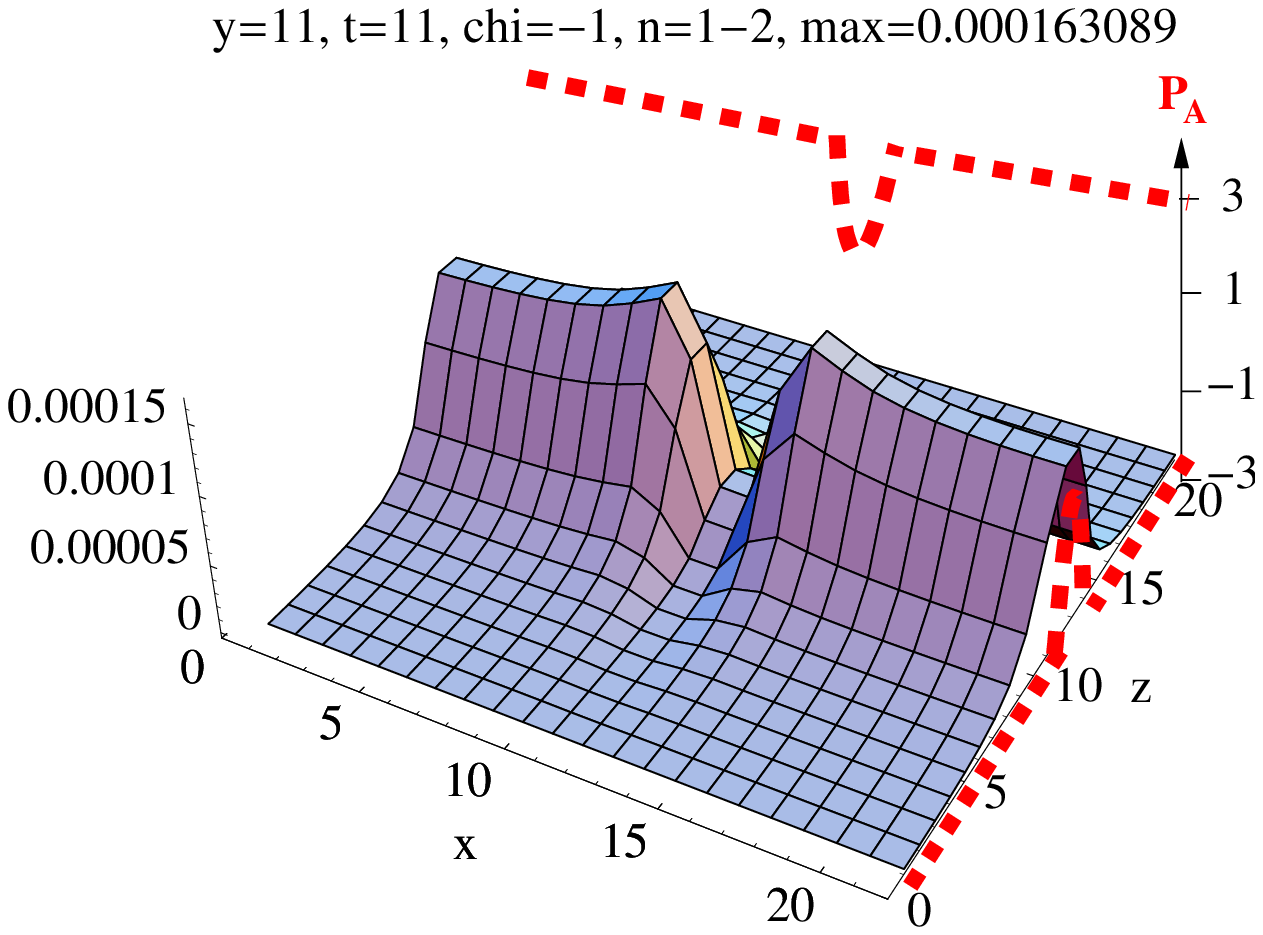}
\textcolor{white}{emty line}\\
e)\includegraphics[width=.44\linewidth]{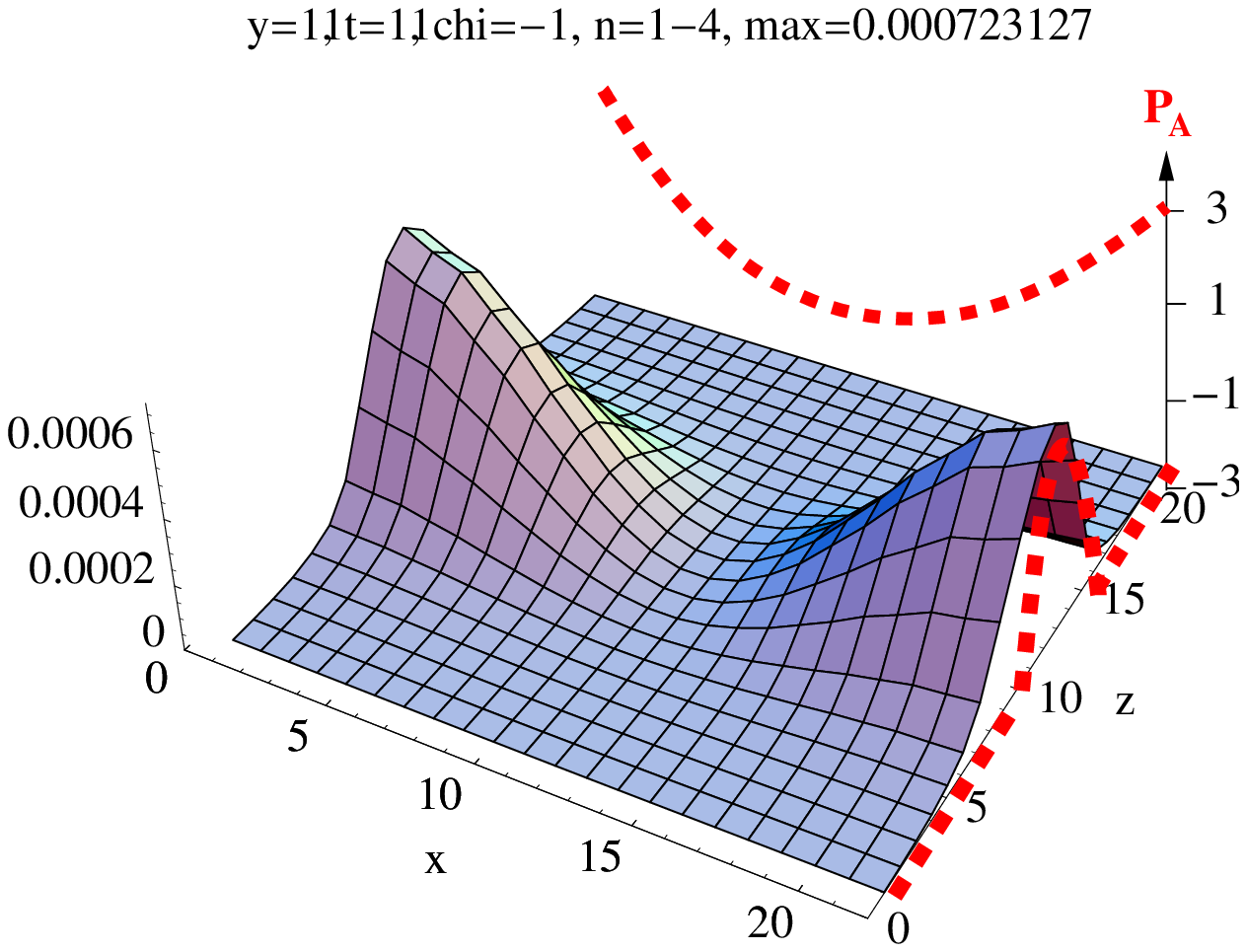}f)\includegraphics[width=.44\linewidth]{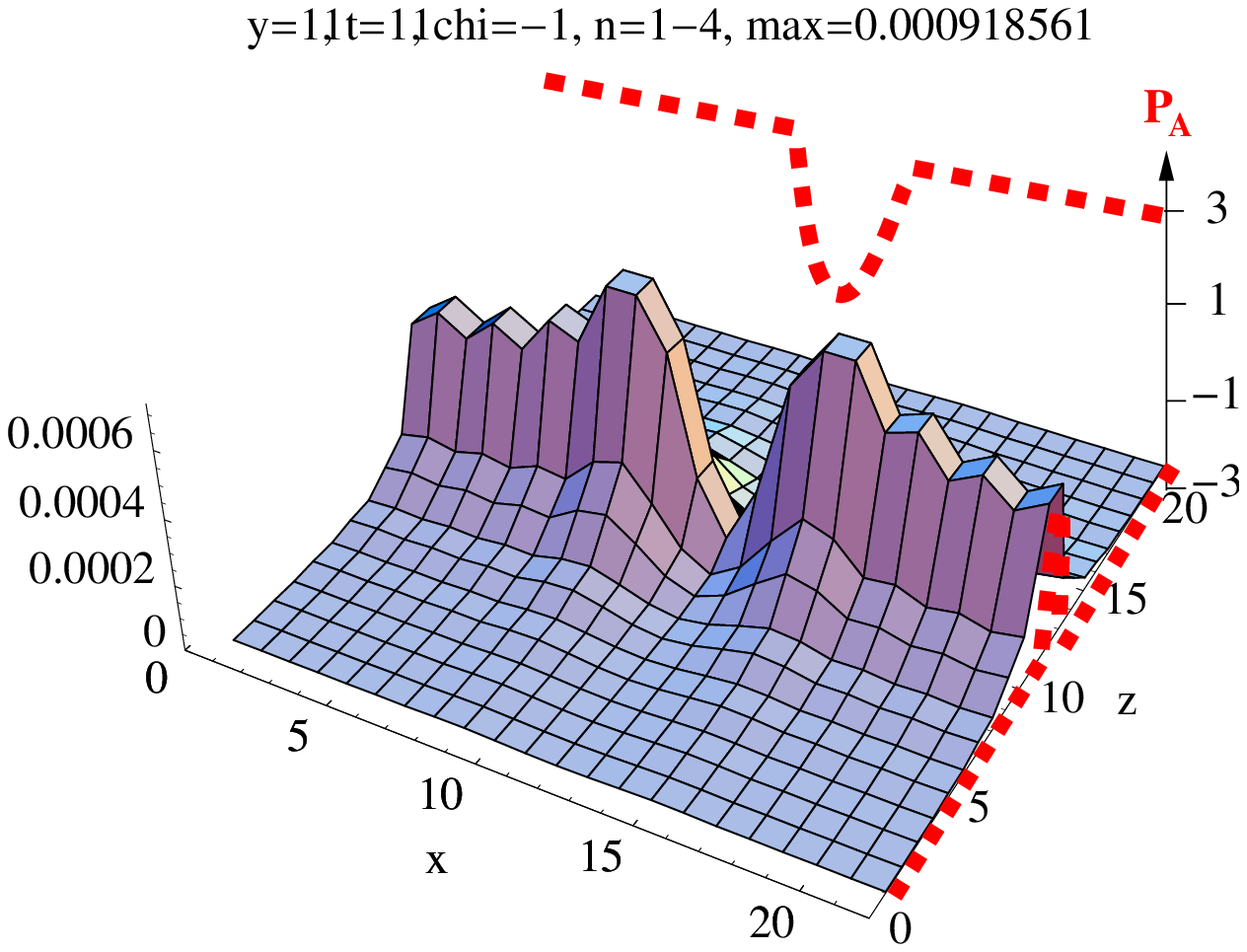}
\caption{"$Q=1/2$ configuration": Scalar eigenmode density of two adjoint overlap (n=1-2) or four adjoint asqtad staggered (n=1-4) zeromodes of negative chirality in the intersection plane of $zt$- and $xy$-vortices for various vortex thicknesses: a) $d_x=d_z=16$ b) $d_x=d_z=12$ c) $d_x=d_z=8$ d) $d_x=d_z=4$ d) $d_x=20$, $d_z=8$ f) $d_x=8$, $d_z=2$. The adjoint zeromodes approach the thick vortex intersection at $x=z=11$ from regions of positive, adjoint Polyakov lines ($P_A$), but do not localize exactly the topological charge contribution $Q=1/2$ because they strictly avoid regions of negative, adjoint Polyakov (Wilson) lines. We also plot the profile of the adjoint Polyakov (Wilson) lines ($P_A$, red dots).}
\label{fig:1hlfmodes2}
\end{figure}

We further vary the vortex thickness $d$, see Fig.~\ref{fig:1hlfmodes2} and do find analogue results: The adjoint zeromodes approach the thick vortex intersection from regions of positive Polyakov lines, but do not localize exactly at the region with topological charge contribution $|Q|=1/2$. They rather spread over the whole lattice avoiding regions of negative traces of adjoint Wilson lines.

\section{Dirac modes in the background of single center vortex pairs}

Since we found that the Dirac zeromodes seem to be more sensitive to the
Polyakov (Wilson) lines than to topological charge contributions we analyze
configurations of single vortex pairs apparently without topological charge. We
use parallel $xy$-vortices with $t$-links in one time-slice ($t_0$) varying in
$z$-direction according to Eq.~\ref{eq:phi-pl0}, shown in
Fig.~\ref{fig:phis}a). Due to the translation symmetry in $x$ and $y$, {\it
i.e.}, parallel to the vortex planes, and the fact that the links vary only
within an abelian U(1)-subgroup of $SU(2)$, our configuration corresponds to an abelian 2-dimensional problem which we can compare with the abelian center vortices on $\mathbbm T^2$ considered in~\cite{Reinhardt:2002cm}. The nontrivial transition functions used in~\cite{Reinhardt:2002cm} are incorporated in our nontrivial links and the analytical result should be equivalent to our configuration with periodic fermion boundary conditions. 

The occurrence of zeromodes can be related to a U(1) index theorem for $\mathbbm T^2$, Ref.~\cite{Reinhardt:2002cm} finds for $m_+$ vortices and $m_-$ anti-vortices a number $\Delta m/2 = |m_+ - m_-|/2$ of zeromodes (for more details see also~\cite{Falomir:1996as}). 
Since we work in four dimensions and with $SU(2)$, we expect to find four times more zeromodes: we have two times more color indices and two times more spinor components. This is indeed our result for periodic boundary conditions in the $x$- or $y$-directions and antiperiodic boundary conditions in the $z$- or $t$-directions, {\it i.e.}, perpendicular to the vortex surfaces~\footnote{For antiperiodic boundary conditions in the $x$- or $y$-directions, {\it i.e.}, parallel to the vortex surfaces we do not find zeromodes. Because of the translation invariance in $x$ and $y$, the Dirac modes should contain a free wave factor with a wave vector $k=(k_x, k_y)$. But for antiperiodic boundary conditions the smallest allowed modulus of $k$ is $\frac{\pi}{aN} \neq 0$ and the corresponding lowest eigenvalue must be greater than zero.}. 

\begin{figure}[p]
\centering
\psfrag{x}{$x$}
\psfrag{y}{$y$}
\psfrag{z}{$z$}
\psfrag{t}{$t$}
a)\includegraphics[width=.44\linewidth]{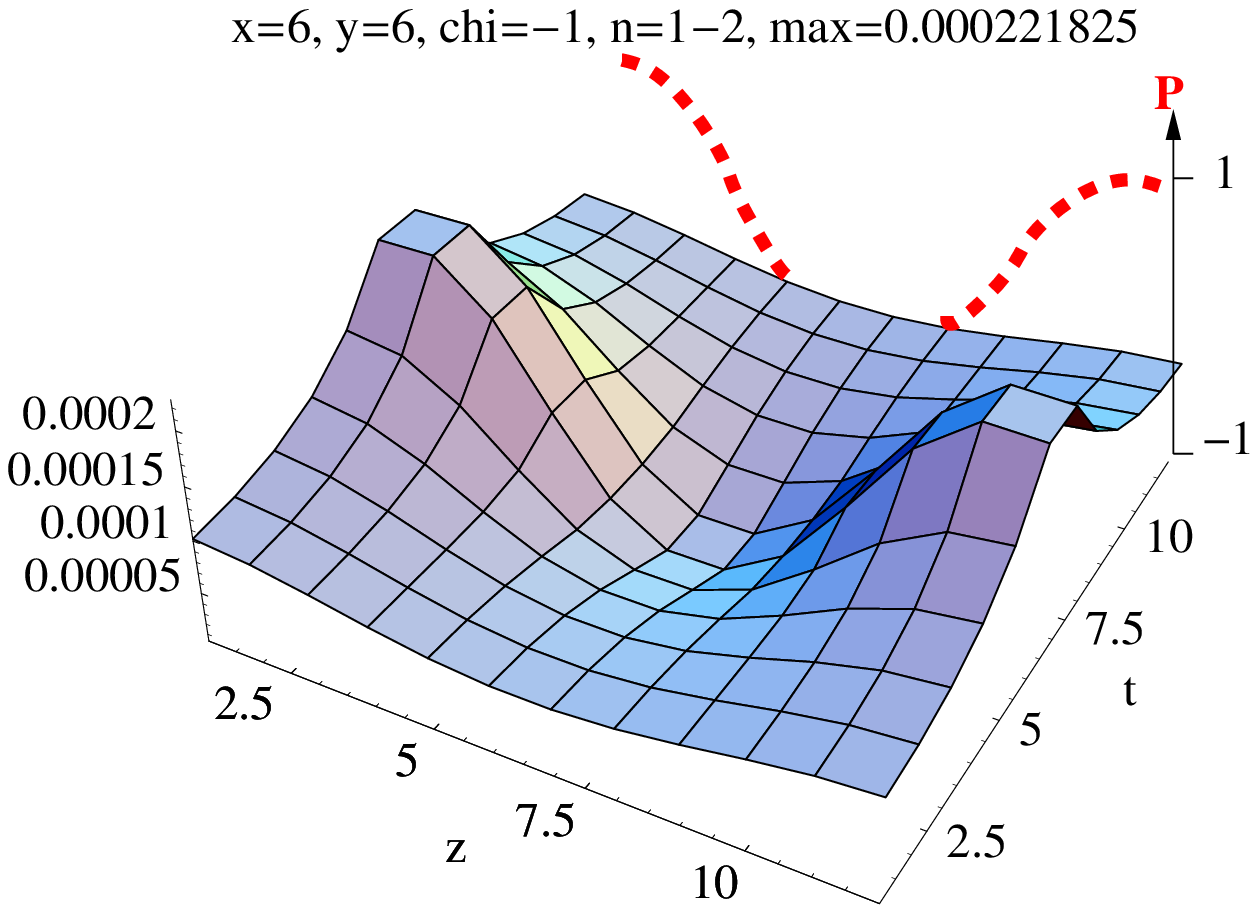}b)\includegraphics[width=.46\linewidth]{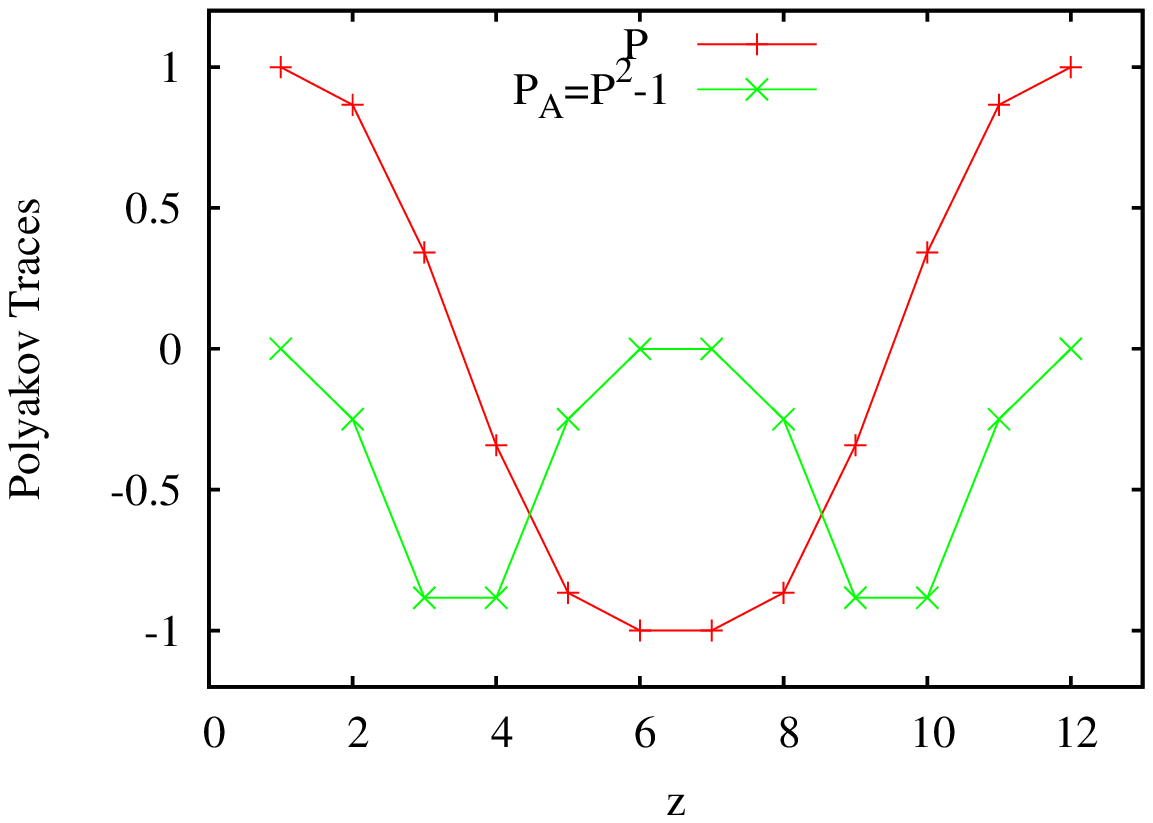}
\textcolor{white}{emty line}\\
c)\includegraphics[width=.46\linewidth]{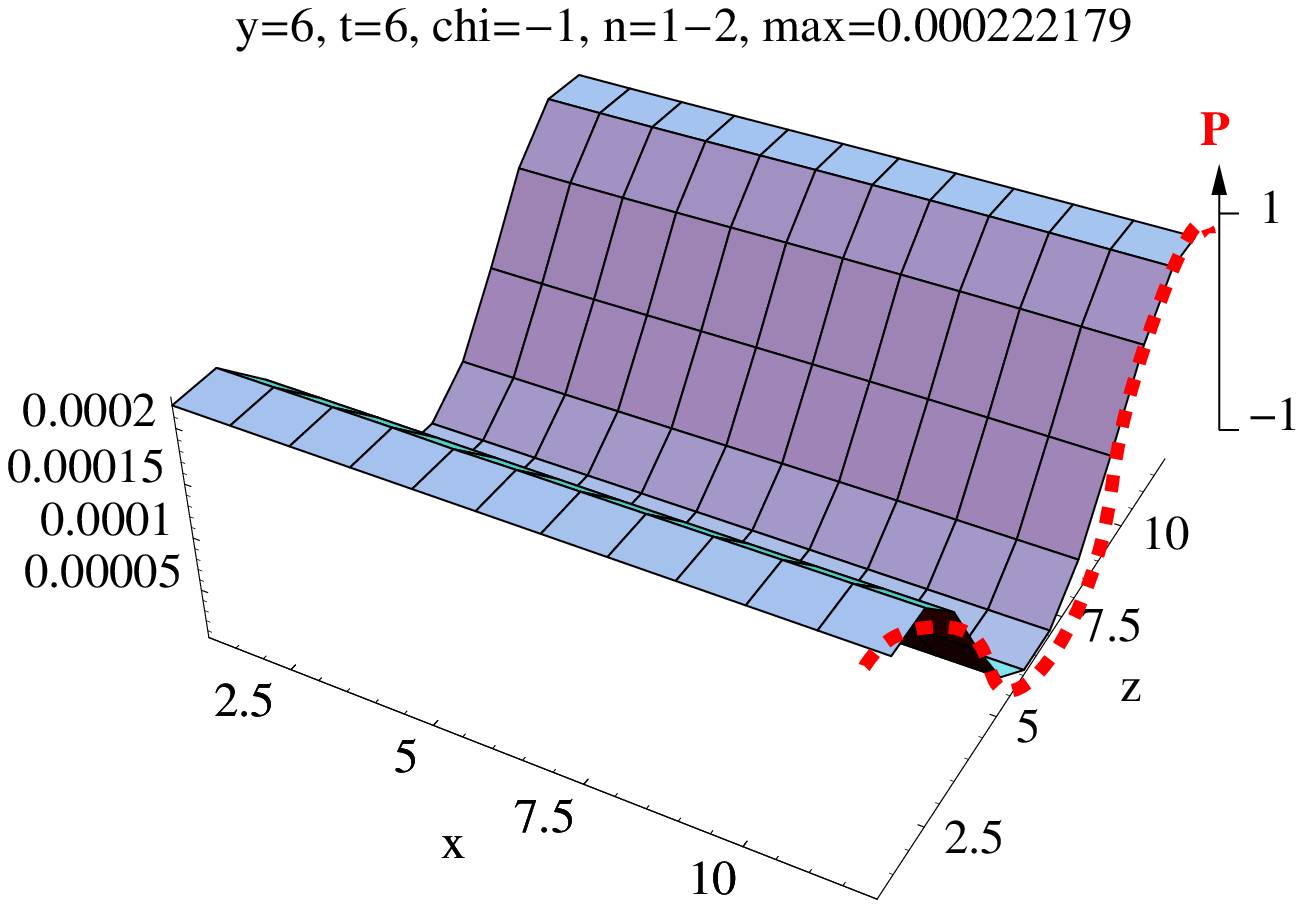}d)\includegraphics[width=.44\linewidth]{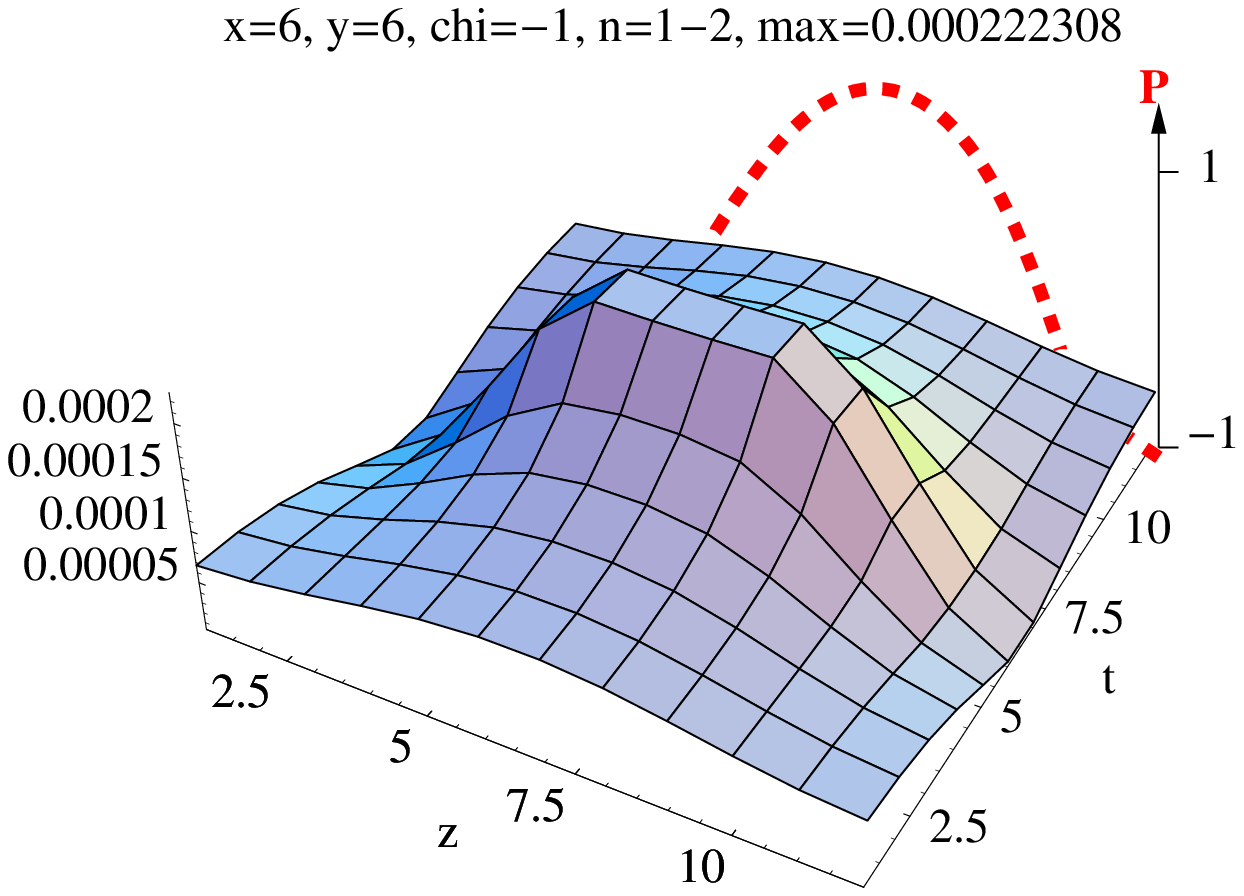}
\textcolor{white}{emty line}\\
e)\includegraphics[width=.44\linewidth]{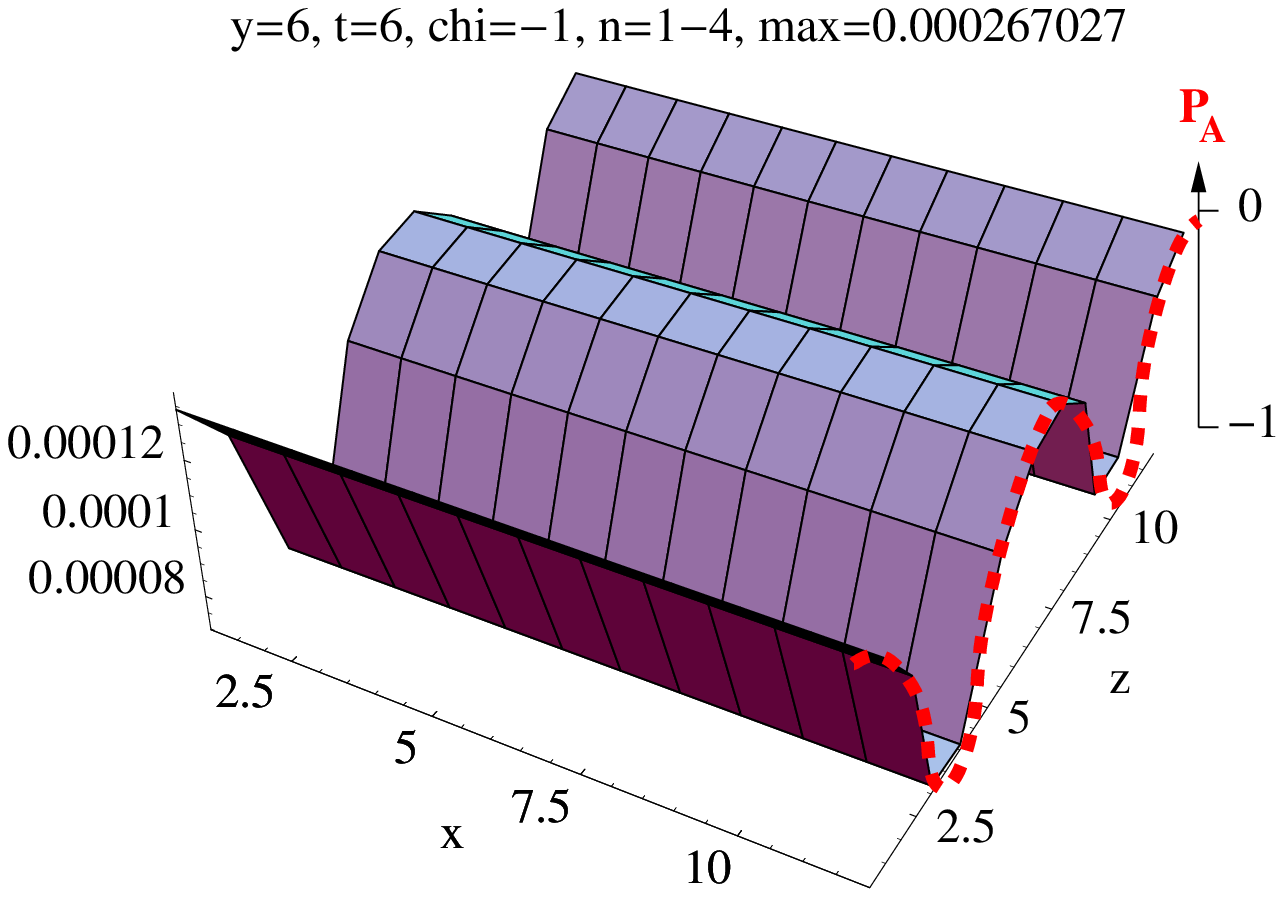}\; f)\includegraphics[width=.48\linewidth]{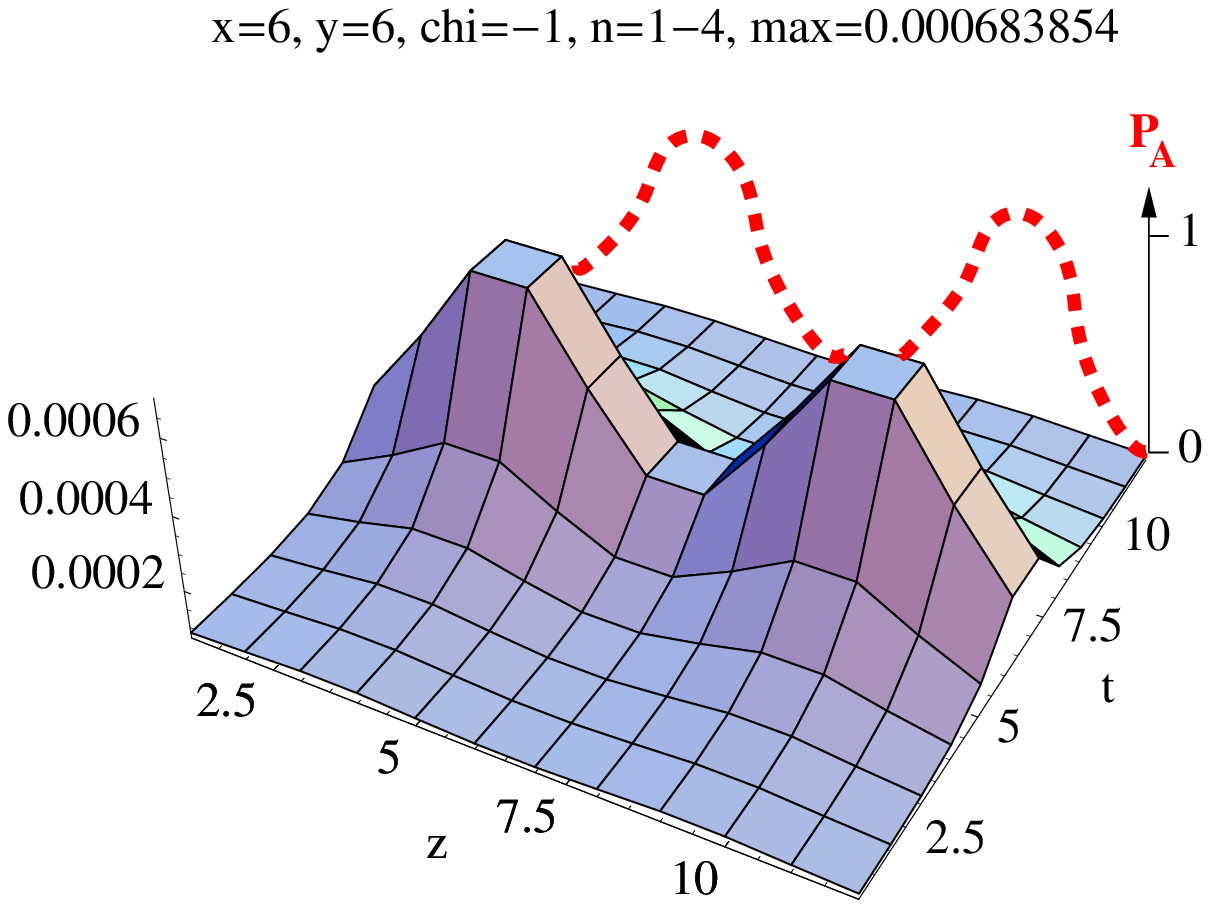}
\caption{Scalar density plots of Dirac zeromodes in the background of a single $xy$-vortex pair at $t=6$ and $z_{1,2}=3.5,9.5$ (see Eq.~\ref{eq:phi-pl0}): a,c) two fundamental zeromodes for each chirality (the four individual modes have equivalent density distributions) show the Polyakov profile at $t=6$ and are constant in $x$- ($y$-) direction; b) profile of the (adjoint) Polyakov traces $P$ ($P_A$) for our configuration (indicated with red dots in the density plots); d) two fundamental zeromodes for each chirality with antiperiodic boundary conditions show the inverted Polyakov profile at $t=6$; e,f) four adjoint zeromodes for each chirality (the eight individual modes again have equivalent density distributions) with periodic resp. antiperiodic boundary conditions.}
\label{fig:pl2modes}
\end{figure}

We find two overlap zeromodes of each chirality\footnote{With the introduction of a little noise, these zeromodes would mix and be lifted to low lying modes.} spreading over the whole lattice parallel to the vortex surfaces, {\it i.e.}, their scalar density distribution is constant in the $x$- and $y$-direction, strictly avoiding regions with negative Polyakov values. At the time-slice with nontrivial $t$-links ($t_0$) the density distribution clearly follows the profile of these links, which correspond to the Polyakov loops. We confirm these results also for staggered fermions and adjoint representations with adjoint Polyakov traces $P_A=P^2-1$ (see Eq.~\ref{eq:adjtrace}). Fig.~\ref{fig:pl2modes} shows our Polyakov profiles $P$ and $P_A$ together with scalar density plots of the fundamental and adjoint Dirac zeromodes with periodic and antiperiodic boundary conditions in $t$-direction (the latter invert the Polyakov profiles, corresponding to a multiplication with $-1$). The individual zeromodes for both chiralities show equivalent density distributions, but they differ from the analytic results found in~\cite{Reinhardt:2002cm}. 
Further, the lowest non-zeromodes also propagate parallel to the vortex surfaces, showing similar density distributions but with opposite response to the Polyakov profiles, {\it i.e.}, exchanged boundary conditions.

The above configuration consists of two identical flux lines because the field strength is invariant under a $z$-translation $T_z$ by half the lattice size, which exchanges the two vortices.
But this symmetry is spoilt by the fact that in the $zt$-plane we need to distinguish two regions which differ by the value of the Polyakov loop $P$.
More generally, we could multiply all links $U_t$ at a fixed $t$-coordinate
(not necessarily $t_0$) by an arbitrary $SU(2)$-element. While this would not alter any plaquettes ({\it i.e.}, the flux), $P$ would change at a given point $z$ and could be modified to any desired value. Invariably, however, the Polyakov loop varies by a center element at the locations of the vortices. This simply follows from Stokes theorem. Two Polyakov loops on either side of the vortex can be deformed into a single curve encircling it. The line integral along this curve equals the center element of the vortex in its interior.

Consequently, $T_z$ corresponds to a center transformation switching the sign of $P$. Since the Polyakov loop is a gauge-invariant quantity, the configurations before and after a translation with $T_z$ cannot be gauge-equivalent and in general the Dirac operator will not be invariant under $T_z$. Since the scalar densities of zeromodes and the near-zeromodes are asymmetric with respect to the translation $T_z$ and the only effect of $T_z$ is a change of $P$, we are forced to conclude that the Dirac eigenmodes are sensitive to the value of the Polyakov loop.

For completeness we mention the anti-parallel vortex (Fig.~\ref{fig:phis}b). Since the total flux vanishes, in accord with the 2-dimensional U(1) index theorem~\cite{Falomir:1996as} we observe no zeromodes. In this case the translation $T_z$ changes both, the direction of fluxes and the value of $P$. Still, the lowest-lying eigenmodes are sensitive to the value of the Polyakov loop with respect to boundary conditions.

\section{Conclusions}
In this paper we have studied low lying asqtad staggered and overlap Dirac operator
eigenmodes in the background of intersecting vortex configurations.
Each intersection region carries topological charge $\pm 1/2$. We
found that the zeromodes are sensitive to the value of the Polyakov
loops (Wilson lines), avoiding regions with negative Polyakov loops.
In particular, we studied configurations with two parallel planar
vortices, which can be exchanged by a discrete translation $T$ by half
the lattice size. While the field strength (and the topological
charge distribution obtained from the field strength) is invariant
under this translation, the Dirac operator is not, since the translation
modifies (shifts) the Polyakov loop. Therefore, the distribution of
the scalar density of the Dirac operator zeromodes is not forced to be
invariant under this translation, and we do find that they are not.
This is in contrast to the analytical results of Ref.~\cite{Reinhardt:2002cm} which appear to be invariant under the discrete translations. This happens even though our planar vortex configurations are presumed to be equivalent to the abelian gauge configurations studied there. We think the difference might be related to differences in the Polyakov loop. 
Configurations with the same flux distribution can differ in the
distribution of the Polyakov loop values, which can lead to differences
in the scalar density of Dirac operator zeromodes, as those tend
to avoid regions of negative Polyakov loops. We find that the zeromodes
in the fundamental representation are not quite spread along the
center vortex sheets and don't quite peak at the vortex intersection
points. Rather, they approach these topological structures from
regions with positive Polyakov lines while avoiding regions of
negative Polyakov lines.

With adjoint fermions we tried to find zeromodes which identify exactly one topological charge contribution $Q=1/2$ of a single vortex intersection. We analyzed linear combinations of zeromodes which maximize the inverse participation ratio (IPR), {\it i.e.}, localize as much as possible. We found that the scalar eigenmode density of adjoint fermions always peaks at least at two intersections. 
Further, we analyzed a lattice configuration with only one "thick" vortex
intersection. Since both, gluonic and (adjoint) fermionic definitions of
topological charge fail to detect intersections with at least one thin
vortex, both definitions of $Q$ merely signal the value $Q=1/2$. We find it
remarkable that the two corresponding adjoint zeromodes are not localized to the
region with nonvanishing topological charge contribution but spread over the
whole lattice, avoiding regions of negative traces of adjoint Polyakov (Wilson) lines. Therefore we conclude that the Dirac zeromodes are more sensitive to the Polyakov (Wilson) lines than to the topological charge contributions. 

Finally we confirm this conjecture with Dirac eigenmodes for configurations
with nontrivial Polyakov profiles of single center vortex pairs, apparently
without topological charge. In principle, the sensitivity of the Dirac
operator to the Polyakov (Wilson) lines with respect to boundary conditions is nothing new. In calorons for example {\it the zeromode hops with the
boundary conditions in the compact
direction}~\cite{GarciaPerez:1999ux,Bruckmann:2007ru} between the monopole constituents, where the Polyakov loop actually passes
through $\mathbbm{1}$ and $-\mathbbm{1}$. 
Since calorons seem only to be important in the deconfinement phase, {\it i.e.}
with finite temporal lattice size, these effects may be interpreted as
finite size effects. Similarly, our specially constructed vortex
configurations discussed in this paper, lacking a natural size for a
correlation length, could be viewed as having large finite size effects.
Thus our observations may not completely hold in a realistic vortex vacuum, where the correlations contained in vortices and Dirac zeromodes decay within a finite length, which should ideally be much smaller than the extent of the lattice. Nevertheless, a better understanding of the sensitivity of Dirac operator eigenmodes to Polyakov (Wilson) lines in realistic situations 
should be pursued, and might shed more light on the mechanism of chiral
symmetry breaking. We plan to study this in future work.

\acknowledgments{We thank Rob Pisarski and Jeff Greensite for the suggestion to investigate fractional topological charge configurations with adjoint fermions. We are grateful to Lorenz von Smekal and Falk Bruckmann for helpful discussions. This research was partially supported by the Austrian Science Fund (``Fonds zur F\"orderung der Wissenschaften'', FWF) under contract P22270-N16 (R.H.).}


\bibliographystyle{unsrt}
\bibliography{../literatur}

\end{document}